\documentclass[twocolumn,secnumarabic,amssymb, nobibnotes,superscriptaddress,showkeys, aps, prd]{revtex4-2}

\usepackage{graphicx}	% Including figure files
\usepackage{amsmath}	% Advanced maths commands
\usepackage{amssymb}	% Extra maths symbols
\usepackage{bm}		% Bold maths symbols, including upright Greek
\usepackage{hyperref}
\usepackage{subfigure}
\usepackage{romannum}

\newcommand{\diff}{{\text{d}}}

\newcommand{\aap}{    {\it Astron. Astrophys.}}

\newcommand{\apjl}{   {\it Astrophys. J. Lett.}}

\newcommand{\mnras}{  {\it Mon. Not. Roy. Astron. Soc.}}

\newcommand{\ssr}{    {\it Space Sci. Rev.}} 
\newcommand{\na}{    {\it New Astronomy}} 
\newcommand{\araa}{ {\it Annual Review of Astronomy and Astrophysics}}
\newcommand{\apjs}{ {\it The Astrophysical Journal Supplement Series}}

\begin{document}

\title[Spectral index Versus luminosity in TDEs]{Disk-corona modeling for spectral index and luminosity correlation of tidal disruption events}%

\author{T. Mageshwaran}
\affiliation{Department of Space Science and Astronomy, Chungbuk National University, 12 Gaeshin-dong, Heungduk-gu, Cheongju 361-763, Korea}
%\affiliation{Department of Astronomy and Astrophysics, Tata Institute of Fundamental Research, Mumbai 400005, India}
\author{Sudip Bhattacharyya}
\affiliation{Department of Astronomy and Astrophysics, Tata Institute of Fundamental Research, Mumbai 400005, India}
\email[Mageshwaran T: ]{tmageshwaran2013@gmail.com}
\date{\today}%

\begin{abstract}
We present a relativistic disk-corona model for a steady state advective accretion disk to explain the UV to X-ray spectral index $\alpha_{\text{OX}}$ evolution of four tidal disruption event (TDE) sources XMMSL2J1446, XMMSL1J1404, XMMSL1J0740, and AT2018fyk. The viscous stress in our model depends on gas ($P_g$) and total ($P_t$) pressures as $\tau_{r\phi} \propto P_g^{\mu} P_t^{1-\mu}$, where $\mu$ is a constant. We compare various steady and time-dependent sub-Eddington TDE accretion models along with our disk-corona model to the observed $\alpha_{\text{OX}}$ of TDE sources and find that the disk-corona model agrees with the observations better than the other models. We find that $\mu$ is much smaller than unity for TDE sources XMMSL2J1446, XMMSL1J1404 and XMMSL1J0740. We also compare the relativistic model with a non-relativistic disk-corona model. The relativistic accretion dynamics reduce the spectral index relative to the non-relativistic accretion by increasing the energy transport to the corona. We estimate the mass accretion rate for all the sources and find that the observed luminosity follows a nearly linear relation with the mass accretion rate. The ratio of X-ray luminosity from the disk to the corona increases with the mass accretion rate. The observed $\alpha_{\text{OX}}$ shows positive and negative correlations with luminosity. The disk-corona model explains the negative correlation seen in the TDE sources XMMSL1J0740, XMMSL2J1446, and XMMSL1J1404. However, TDE AT2018fyk shows a positive correlation at higher luminosity and shows a better fit when a simple spherical adiabatic outflow model is added to the relativistic disk-corona model. Even though the disk luminosity dominates at a higher mass accretion rate, we show that the accretion models without a corona are unable to explain the observations, and the presence of a corona is essential. 
 \end{abstract}

%\keywords{accretion, accretion disks; black hole physics; radiation: dynamics; transients: tidal disruption events (XMMSL1J0740, XMMSL2J1446, XMMSL1J1404, AT2018fyk)}

\maketitle
%\tableofcontents

\section{Introduction}

Black holes have been observed at various scales ranging from stellar-mass black holes to supermassive black holes at the galactic center. The black holes accrete the surrounding gas through viscous accretion and emit radiation in various spectral wavelengths \citep{2002apa..book.....F}. Tidal disruption events (TDEs) occur when a star is disrupted by the gravitational tidal forces of supermassive black holes  when the stellar orbital pericenter $ r_p \leq r_t $, where the tidal radius depends on black hole mass $M_{\bullet}$ and stellar mass $M_{\star}$ and radius $R_{\star}$ as $r_t \simeq (M_{\bullet}/M_{\star})^{1/3} R_{\star}$ \citep{1988Natur.333..523R}. The disrupted debris forms an accretion disk through stream-stream interactions \citep{2016MNRAS.455.2253B,2020A&A...642A.111C} and is viscously accreted resulting in the emission at various wavelengths. The mass accretion rate shows extreme variations ranging from super-Eddington to sub-Eddington and thus constitutes an ideal lab to study the accretion state transitions. The detection of long-term optical and UV emissions ($\sim$ few years) implies that the TDE disks are relatively stable. The late-time X-ray observations for various TDEs by \citet{2020ApJ...889..166J} exhibit diverse X-ray behaviour at late times. They hypothesized that the marked spectral differences may be due to a late-time state change and the X-ray bright phase is often delayed with respect to the UV/optical peak. 

The steady and time-dependent accretion models have been proposed for TDEs at various super and sub-Eddington phases \citep{2020SSRv..216..114R}. The sub-Eddington disk with pressure dominated by gas pressure has been modeled both analytically using self-similar formulation \citep{1990ApJ...351...38C} and numerically \citep{2019MNRAS.489..132M}. A relativistic thin disk model with a fallback at the outer radius was constructed by \citet{2020MNRAS.496.1784M} and they showed that the late time luminosity decline is higher than the luminosity obtained assuming $L \propto \dot{M}_{\text{fb}}$, where $\dot{M}_{\text{fb}}$ is the mass fallback rate of the disrupted debris. The super-Eddington disk is more complex to study due to the presence of strong radiation pressure that results in an outflow. These super-Eddington disk has been modeled using a steady slim disk accretion model with an adiabatic and spherical outflow \citep{2009MNRAS.400.2070S}, whereas a time-dependent and self-similar model for a non-radiative disk was developed by \citet{2014ApJ...784...87S}. A time-dependent and self-similar model with mass infall to the disk was constructed by \citet{2021NewA...8301491M} for both sub and super-Eddington disk with outflow. These models are used to fit the optical/UV observations. 

The X-ray spectrum often shows a power law decline with frequency instead of a blackbody decline at higher frequencies.
The excess X-ray emission in both stellar mass black holes and active galactic nuclei is explained by including a corona surrounding the accretion disk \citep{2006ARA&A..44...49R,2012MNRAS.420.1848D}. The geometry of an accretion disk depends on the mass accretion rate normalized to the Eddington rate given by $\dot{m}$ and the pressure in the disk. The disks with high $\dot{m}$ are thick disks with pressure dominated by radiation, whereas the disks with low $\dot{m}$ are thin disks with pressure dominated by gas pressure. The observations have revealed that the AGN show some characteristics of X-ray binaries and these similarities often suggests that the AGN behave like X-ray binaries that are scaled up in size \citep{2019ApJ...883...76R}. The thermal emission peaks in UV for AGN disks whereas it dominates in soft X-ray for X-ray binaries. The dynamics of accretion flow around black holes of mass at various scales are compared using the UV to X-ray spectral index ($\alpha_{\text{OX}}$) between 2500 \AA ~and 2 keV. \citet{2020MNRAS.497L...1W} showed the correlation between $\alpha_{\text{OX}}$ and luminosity for seven TDEs and obtained a statistically significant empirical correlation through fit. They found on average that the sources with higher Eddington luminosity ratio have disk-dominated X-ray spectra, with high $\alpha_{\text{OX}}$ values and a small power-law contribution to the total X-ray flux. Sources at lower Eddington ratio have power-law dominated X-ray spectra and lower $\alpha_{\text{OX}}$ values. 

In this paper, we propose an advective disk-corona emission model for a relativistic disk. We use the steady accretion model with energy loss to the corona and include the gravitational and Doppler redshift for a relativistic case. We use a viscosity similar to an alpha viscosity and is a combination of total pressure and gas pressure. The ratio of tidal radius ($r_t$) to black hole horizon ($r_H$) is given by $r_t/r_H \propto M_{\bullet}^{-2/3}$ and thus the relativistic disk-coronal model is crucial for TDEs with higher black hole mass. The relativistic dynamics are also crucial for accretion near the inner radii. We present the importance of corona in explaining the observed spectral index $\alpha_{\text{OX}}$ and will compare the disk-corona emission model with the other TDE accretion model to the observed spectral index for four sources XMMSL2J1446, XMMSL1J1404, XMMSL1J0740 and AT2018fyk. We estimate the physical parameters of the models by comparing them to the observations and evaluate the mass accretion rate for each observation of the sources. We have also developed a non-relativistic advective disk-corona model shown in the appendix for comparison with the relativistic model. Here, we propose the relativistic model as the primary model for the observational fitting and the non-relativistic model is for showing the impact of relativistic dynamics on the estimation of the parameters. Even though both non-relativistic and relativistic models fit the observations, the obtained parameters show significant variations.

In section \ref{rdcm}, we present the relativistic formulation of the disk-corona accretion and their emissions. We discuss the viscous stress, coronal structure, and disk-corona emission. The non-relativistic model is presented in appendix \ref{ndcm}. We discuss the spectral index and its formulation in section \ref{aox}. We compare the various steady and time-dependent accretion models without corona to the observations in section \ref{modcomp} and show the importance of the disk-corona model in explaining the observations. We discuss our results in section \ref{discuss} and present the summary in section \ref{summary}. 

\section{\label{rdcm} Relativistic disk-corona model}

The disk-corona model we present here is based on the standard conservation equations of a vertically integrated and optically-thick accretion disk coupled with X-ray corona. A fraction of the energy is transported to the corona most probably by the magnetic fields generated in the disk. The corona above the disk is heated via magnetic reconnections and the corona energy flux at any radius $r$ is given by $Q_{\text{cor}} = v_D P_{\text{mag}}$. The magnetic pressure is given by $P_{\text{mag}} = B^2/(8 \pi)$ with magnetic field $B$ and $v_D$ is the drift velocity taken proportional to the Alfvén speed $v_A$ via an order-unity constant $b$ \citep{2019A&A...628A.135A}. The drift velocity is given by $v_D = b \sqrt{2 P_{\text{mag}}/\rho}$, where $\rho$ is the density. The stress tensor is assumed to be dominated by Maxwell stresses \citep{2015ApJ...808...54M} and thus given by $\tau_{r\phi} = k_0 P_{\text{mag}}$, where $k_0$ is of order unity. For simplicity, we take $k_0$ and $b$ to be unity in our calculations. The numerical simulations have shown that magneto-rotational instability (MRI) growth rate depends on the ratio of radiation to gas pressure \citep{2002ApJ...566..148T} and thus the magnetic pressure is approximated to \citep{2019A&A...628A.135A}

\begin{equation}
P_{\text{mag}} = \alpha_0 P_g^{\mu} P_t^{1-\mu},
\label{ptot}
\end{equation}
 
\noindent with the total pressure $P_t= P_r + P_g$, radiation pressure $P_r = a T^4/3$ where $a$ is a radiation constant, and the gas pressure is given by $P_g =  k_B \rho T/(\mu_m m_p)$, where $k_B$ is the Boltzmann constant, $\mu_m$ is the mean molecular weight taken to be ionized solar mean molecular weight of $0.65$, $m_p$ is the mass of a proton and $T$ is the temperature in the disk. The energy radiated by the disk is Compton scattered in the corona and some fraction of the scattered photons travels downward to the disk where it gets absorbed and some fraction of it is reflected without being absorbed known as disk albedo. We assume the downward component of the X-ray emission to be $\eta$ and a disk albedo to be $a_d$. 

The energy conservation equation of the disk is given by

\begin{equation}
Q^{+}_{\text{v}} - Q_{\text{adv}} - Q_{\text{cor}} + \eta (1-a_d) Q_{\text{cor}} = Q_{\text{rad}},
\label{eqnss}
\end{equation}

\noindent where $Q_{\text{rad}}$ is the radiative energy loss and $Q_{\text{adv}}$ is the advective energy loss which is important if the viscous stress is dominated by total pressure. An anisotropic Comptonization in plane-parallel geometry provides a typical value of $\eta = 0.55$ and $a_d = 0.1$ \citep{1993ApJ...413..507H}. In general, the quantity $\eta$ and $a_d$ are a function of the photon index of the X-ray spectrum from the corona and depend on a detailed radiative transfer but we here assume it to be a constant free parameter.

We use the relativistic accretion disk equations in the Kerr metric given in appendix \ref{armp}. The space-time metric in the geometrical units ($c=G =1$) with the signature ($- + + +$). The vertically integrated disk equations are reduced to a steady solution by taking the time derivative to be zero. The mass conservation equation given by equation (\ref{massr}) results in 

\begin{equation}
\dot{M} = -2 \pi r \Sigma u^r = - 2 \pi \Sigma \frac{V}{\sqrt{1-V^2}} \Delta^{1/2}.
\end{equation}

Assuming a circular motion of the matter in the accretion disk with azimuthal velocity given by 

\begin{equation}
\mathcal{L} = \frac{M^{1/2}(r^2 - 2 a M^{1/2}r^{1/2}+a^2)}{r^{3/4}(r^{3/2}- 3 M r^{1/2} + 2 a M^{1/2})^{1/2}}.
\label{mathl}
\end{equation}

\noindent Using this and solving the angular momentum conservation equation given by equation (\ref{azir}), we have 

\begin{equation}
r \bar{S}^r_{\phi} = \frac{\dot{M}}{2 \pi } (\mathcal{L} - \mathcal{L}_{\text{in}}),
\label{srphi}
\end{equation}

\noindent where $\mathcal{L}_{\text{in}}$ is the angular momentum at the inner radius and we assume that the viscous stress is zero at the inner radius. The $\bar{S}^r_{\phi}$ is given by

\begin{equation}
\bar{S}_{\phi}^r = - \nu \Sigma \frac{\Delta^{1/2} A^{3/2} \gamma_L^3}{r^5 } \frac{\partial \Omega}{\partial r},
\label{sphir}
\end{equation}

\noindent where $\Omega = 2 M a r/A + r^3 \Delta^{1/2} \mathcal{L}/(\gamma_L A^{3/2})$. For $\mathcal{L}$ given by equation (\ref{mathl}), the $\Omega$ reduces to Keplerian velocity $\Omega_K$ given by 

\begin{equation}
\Omega_K = \frac{M^{1/2}}{r^{3/2} + a M^{1/2}}. 
\label{omek}
\end{equation}

The tetrad component of shear stress $S^{\alpha \beta}$ is identical to the viscous stress $-t_{r \phi}$ \citep{1995ApJ...450..508R} and thus the vertically integrated viscous stress $\tau_{r \phi} = r^2/(\gamma_L A^{1/2} \Delta^{1/2}) S^r_{\phi}$. Following viscous stress $\tau_{r\phi}$ discussed in section \ref{ndcm}, and equation (\ref{srphi}), we have

\begin{equation}
k_0 \alpha_0 \frac{P_g^{\mu} P_t^{3/2 - \mu}}{\sqrt{\rho}} = \frac{1}{2} \frac{\dot{M}}{2\pi} \sqrt{\frac{f(x,j)}{x}} \frac{\mathcal{L}-\mathcal{L}_{\text{in}}}{\gamma_L A^{1/2} \Delta^{1/2}},
\label{releq1}
\end{equation} 

\noindent where $x = r/r_g$ with $r_g = G M_{\bullet}/c^2$. We considered the disk height obtained by using equation (\ref{mathl}) in equation (\ref{height}) and is given by $H = \sqrt{r^3/G M_{\bullet}} c_s/\sqrt{f(x,~j)}$, with $f(x,~j)$ given by \citep{2020MNRAS.496.1784M}

\begin{equation}
f(x,~j) = \left[1 - \frac{4 j}{x^{3/2}} + \frac{3 j^2}{x^2}\right] \left[1 - \frac{3}{x} + \frac{2 j}{x^{3/2}}\right]^{-1}.
\end{equation} 

The viscous heating is given by 

\begin{equation}
Q_{\text{v}} = \nu \Sigma \frac{\gamma_L^4 A^2}{r^6} \left(\frac{\partial \Omega}{\partial r}\right)^{2}.
\end{equation}

\noindent In the steady state accretion, the advection energy flux is given by (following \cite{2011ApJS..195....7X})

\begin{equation}
Q_{\text{adv}} = \frac{\dot{M} c_s^2}{2 \pi r^2} \left(\frac{4 - 3\beta_g}{\Gamma_3-1}\right)\left[-\frac{r}{T}\frac{\partial T}{\partial r}+(\Gamma_3-1)\frac{r}{\rho}\frac{\partial \rho}{\partial r}\right],
\end{equation}

\noindent where $\beta_g = P_g/P_t$ is the ratio of gas to total pressure, and   

\begin{equation}
\Gamma_3-1 = \frac{(4 - 3 \beta_g)(\gamma-1)}{\beta_g + 12 (1-\beta_g) (\gamma-1)}
\end{equation}

\noindent and $\gamma$ is the ratio of specific heats for constant pressure to constant volume \citep{1939isss.book.....C}. The radiative flux is given by $Q_{\text{rad}} = 4 \sigma T^4 / (3 \tau)$, where $T$ is the disk mid plane temperature and the opacity $\tau = \tau_{\text{es}} + \tau_{a} = (\kappa_{\text{es}} + \kappa_a)\Sigma$, is the sum of Thomson opacity due to electron scattering and Kramers' opacity due to absorption. The Thomson opacity is given by $\kappa_{\text{es}}=0.34~{\rm cm^{2}~g^{-1}}$, and the Kramers' opacity is given by $\kappa_a = \kappa_0 \rho T^{-7/2} ~{\rm cm^{2}~g^{-1}}$, where $\kappa_0 = 2 \times 10^{24}$ assuming $\rho$ and $T$ are in the cgs unit \citep{2020ApJ...894....2P}. Thus, the energy conservation equation given by equation (\ref{eqnss}) is 

\begin{multline}
\frac{\dot{M} c_s^2}{2 \pi r^2} \frac{4 - 3\beta_g}{\Gamma_3 - 1}\left[ -\frac{r}{T} \frac{\partial T}{\partial r} + \left(\Gamma_3 - 1\right) \frac{r}{\rho} \frac{\partial \rho}{\partial r} \right] = \\ Q_{\text{v}}- Q_{\text{cor}} + \eta (1- a_d)Q_{\text{cor}} - Q_{\text{rad}},
\label{releq2}
\end{multline}

\noindent where the energy flux transported to corona is $Q_{\text{cor}} = v_D P_{\text{mag}}$. We obtain the solution of density and temperature by solving equations (\ref{releq1}) and (\ref{releq2}) with density and temperature zero at the inner radius taken to be the innermost stable circular orbit (ISCO). The effective temperature of the disk $T_{\text{eff}} = (Q_{\text{rad}}/\sigma)^{1/4}$ and thus, the luminosity in various spectral bands following a blackbody emission. The flux density of the disk radiation, as seen by a distant observer at rest, is given by

\begin{equation}
F_{\nu} (\nu_{\text{obs}}) = \int I_{\nu}(\nu_{\text{obs}}) \, \diff \Theta, 
\end{equation}

\noindent where $\diff \Theta$ is the differential element of solid angle subtended at the observer’s sky by the disk element and $I_{\nu}(\nu_{\text{obs}})$ is the intensity at the observer's wavelength. We approximate the differential element in the Newtonian limit given by 

\begin{equation}
\diff \Theta = \frac{\diff \mathcal{A} }{D_L^2} \cos\theta_{\text{obs}},
\label{dtheta}
\end{equation}

\noindent where $D_L$ is the luminosity distance of the source to the observer, $\diff \mathcal{A}$ is the area element of the disk and $\theta_{\text{obs}}$ is the viewing angle of the observer to the disk. The area of disk in $\{r,~\phi\}$ plane is given by $\displaystyle{\diff \mathcal{A} = \sqrt{A/\Delta}~ \diff r \diff \phi}$ \citep{2020MNRAS.496.1784M}. The gravitational redshift effect is included by using the Lorentz invariant $I_{\nu}/\nu^3$ \citep{1979rpa..book.....R}, such that $I_{\nu}(\nu_{\text{obs}}) = g^3 I_{\nu}(\nu_{\text{em}})$, where $\nu_{\text{em}}$ is the emitted frequency and $g$ is the redshift factor. By taking in account the gravitational and kinematic redshift effects, the redshift factor $g$ is given by \citep{2019MNRAS.489..132M}

\begin{equation}
g = \left[1 - \frac{3}{x} + \frac{2 j}{x^{3/2}}\right] \left[1 + \frac{j}{x^{3/2}}\right]^{-1}.
\end{equation}  

\noindent Thus, observed flux from the disk is given by 

\begin{equation}
F_{\nu} (\nu_{\text{obs}}) = \frac{\cos\theta_{\text{obs}}}{D_L^2} \int g^3 I_{\nu,{\text{em}}}\left(\frac{\nu_{\text{obs}}}{g}\right) \, \diff \mathcal{A}, 
\end{equation}

In the curved space-time, the optical depth for the photons passing through the corona is calculated using the ray-tracing method. We here relax the detailed ray-tracing of the photons and consider a simplified corona emission by assuming the flux to be $(1-\eta) Q_{\text{cor}}$, where $\eta$ is the downward component. The X-ray flux from the corona is calculated as $\displaystyle{F_{X,~{\text{cor}}} =  \frac{\cos\theta_{\text{obs}}}{D_L^2} \int g^4 (1-\eta) Q_{\text{cor}}\,\diff \mathcal{A}}$. We then assume a power-law X-ray spectrum (similar to the non-relativistic model) and calculate the flux in a given frequency $\nu$ as

\begin{equation}
F_{\nu,~{\text{cor}}} = F_{X,~{\text{cor}}} (2-\Gamma) \frac{\nu^{1-\Gamma}}{\nu_f^{2-\Gamma}-\nu_i^{2-\Gamma}}.
\end{equation}

In the next section, we use the emission in UV and X-ray bands from the disk and corona to calculate the spectral index.

\section{\label{aox} Spectral index $\displaystyle{\alpha_{\text{OX}}}$}

The UV to X-ray spectral index is given by 

\begin{equation}
\alpha_{\text{OX}} = 1- \frac{\log_{10}(\lambda L_{2500 \AA})-\log_{10}(\lambda L_{\text{2 keV}})}{\log_{10}(\nu_{\text{2500 \AA}})-\log_{10}(\nu_{\text{2 keV}})},
\label{aox}
\end{equation}

\noindent where $\lambda L_{2500 \AA}$ and $\lambda L_{\text{2 keV}}$ are the luminosities at 2500 \AA~ and 2 keV. The correlation between the spectral index and bolometric luminosity ($L_{\text{bol}}$) normalized to Eddington luminosity ($L_E$) (also called bolometric Eddington ratio) has been studied earlier for X-ray binaries and AGN. In AGN, the thermal emission from the disk peaks in the UV, while the Comptonized emission dominates the X-rays whereas, in X-ray binaries, the thermal emission peaks in soft X-rays whereas the Comptonized coronal emission dominates the hard X-rays. Thus, in X-ray binaries, the correlation study needs a wavelength higher than 2500 \AA~ and 2 keV for thermal and hard component emissions. In a thin disk, the effective temperature scales as $T_{\text{eff}} \propto M_{\bullet}^{-1/4} \dot{m}^{1/4} (r/r_g)^{-3/4}$, and with an increase in black hole mass, the disk effective temperature decreases and the thermal emission peaks in UV. \citet{2011MNRAS.413.2259S} modeled the observed X-ray spectral evolution of the X-ray binary during an outburst and then scaled the evolving X-ray spectra up to AGN. The spectral index versus Eddington luminosity ratio correlation changes its sign for $\dot{m} \lesssim 1 \% $. The correlations are either positive or negative depending on the state of the accretion flow. \citet{2019ApJ...883...76R} studied the correlation for a large sample of luminous broad-line AGN with different luminosities and found that the spectral behaviour of AGN and X-ray binaries (scaled up to AGN) are similar with a positive correlation for high bolometric Eddington ratio and negative correlation for low bolometric Eddington ratio. The transition in correlation occurs around $L_{\text{bol}}/L_E \sim 10^{-2}$. 

The spectral index for TDEs also shows similar positive and negative correlations and the transition occurs around $L_{\text{bol}}/L_E \sim 10^{-2}$ \citep{2020MNRAS.497L...1W}. They approximated the bolometric luminosity as a sum of X-ray luminosity ($0.01-10~{\text{keV}}$) and the UV luminosity. The X-ray spectrum of the source with a high Eddington ratio is dominated by thermal emission from the disk, whereas the x-ray spectrum of the source with a low Eddington ratio is dominated by non-thermal power-law emission. Such similar results for AGN and X-ray binaries have been shown by \citet{2019ApJ...883...76R}. The timescale of TDE evolutions is very much smaller than the timescale of AGN and thus TDEs are ideal to study the accretion state transitions around supermassive black holes.

Here, we consider the three TDEs XMMSL1J0740 \citep{2017A&A...598A..29S}, XMMSL2J1446 \citep{2019A&A...630A..98S} and XMMSL1J1404 \citep{2020MNRAS.497L...1W}, which show negative correlations implying the X-ray spectrum is dominated by power-law and suggest the Comptonization of the radiation from the disk by the corona. These sources are ideal to test the disk-corona model. Following \citet{2020MNRAS.497L...1W}, we assume the bolometric luminosity as a sum of X-ray luminosity ($0.01-10~{\text{keV}}$) and the UV luminosity (1000-4000 \AA). We consider the X-ray luminosity from both the disk and corona whereas the UV luminosity from the disk only. We take the black hole mass and the spectral index from the literature and are given in Table \ref{parset}. We compare the disk-corona model to the observed $\alpha_{\text{OX}}$ versus luminosity curve to estimate the parameters of the disk and corona.   

\begin{table}
\caption{\label{parset} We summarize the parameters for the considered TDE sources given in \citet{2020MNRAS.497L...1W}. We take the mean value in our calculations.}
\begin{ruledtabular}
\begin{tabular}{cccc}
&&&\\
Sources & Photon index & $M_{\bullet}$ & $D_L$   \\
&&&\\
& ($\Gamma$) & [$\log_{10}(M_{\odot})$] & (Mpc) \\
&&& \\
\hline
&&&\\
XMMSL1J0740 & 1.95 $\pm$ 0.29 & 7.05 $\pm$ 0.43 & 75 \\
&&&\\
\hline
&&&\\
XMMSL2J1446 & 2.58 & 7.79 $\pm$ 0.55 & 127 \\
&&&\\
\hline
&&&\\
XMMSL1J1404 & 2.7 $\pm$ 0.33 & 6.71 $\pm$ 0.40 & 190 \\
&&&\\
\end{tabular}
\end{ruledtabular}
\end{table}

\section{\label{modcomp} Models comparison}

In this section, we compare the various steady and time-dependent sub-Eddington accretion models to the observed spectral index for the three sources given in Table \ref{parset}. We first utilize the accretion models without corona given in the literature and show their inadequacy to explain the observations. The models considered here are the standard steady accretion model with alpha viscosity, steady slim disk model with advection, time-dependent self-similar model without fallback, and time-dependent self-similar model with fallback. 

\begin{figure}
\begin{center}
\includegraphics[scale=0.6]{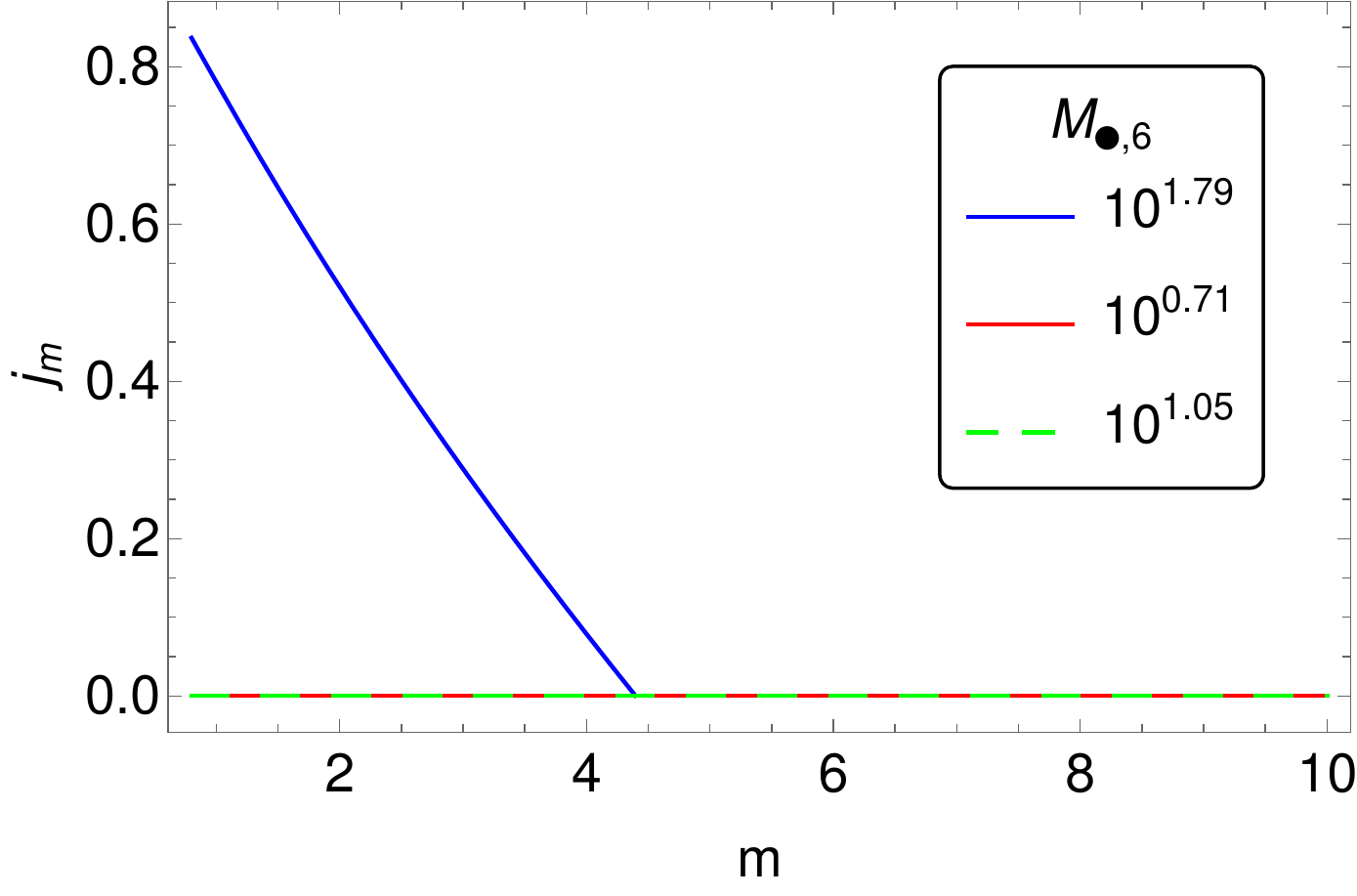}
\end{center}
\caption{\label{jmint} The critical black hole spin $j_m$ required for the tidal radius to be greater than the ISCO radius such that a stable circular accretion disk can be formed. The $j_m$ is calculated for the source given in Table \ref{parset} as a function of stellar mass $m = M_{\star}/M_{\odot}$. The disrupted debris forms a circular accretion disk with an inner radius at ISCO for $j \geq j_m$. The red and green lines have $j_m = 0$. We assume prograde spin only as the stellar orbit is co-rotating with the black hole and the minimum value of $j_m = 0$. See section \ref{modcomp} for details.}
\end{figure}

We normalize the black hole mass as $M_{\bullet,6} = M_{\bullet}/[10^6 M_{\odot}]$ and the stellar mass as $m = M_{\star}/M_{\odot}$ with stellar radius given by $R_{\star} = R_{\odot} m^{0.8}$ \citep{1994sse..book.....K}. We assume that the tidal radius lies above the ISCO radius such that a stable circular accretion disk can be formed and this requires a condition given by $Z(j) \leq r_t/r_g$ and thus a minimum spin value given by $j_{m} = j_m(M_{\bullet,6},~m)$ which is shown in FIG.~\ref{jmint}. We consider the prograde spin only and thus the minimum value of $j_m = 0$. The black hole mass is taken from the Table \ref{parset} and the stellar mass and black hole spin is taken in the range of $m \in \{0.8,~10\} $ and $ j \in \{j_m,~0.9\}$. 

\begin{figure}
\begin{center}
\subfigure[XMMSL2J1446]{\includegraphics[scale=0.55]{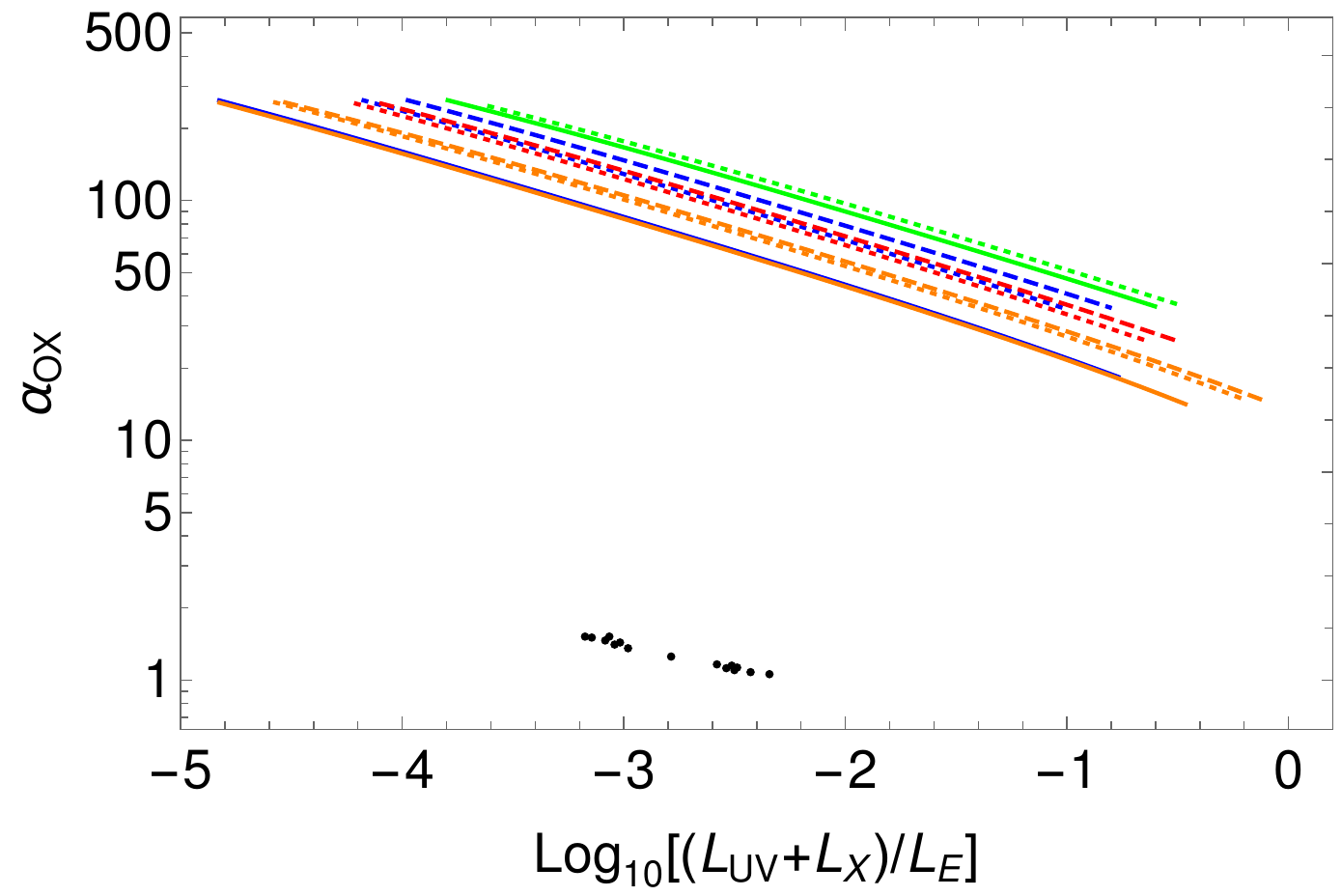}}
\subfigure[XMMSL1J1404]{\includegraphics[scale=0.55]{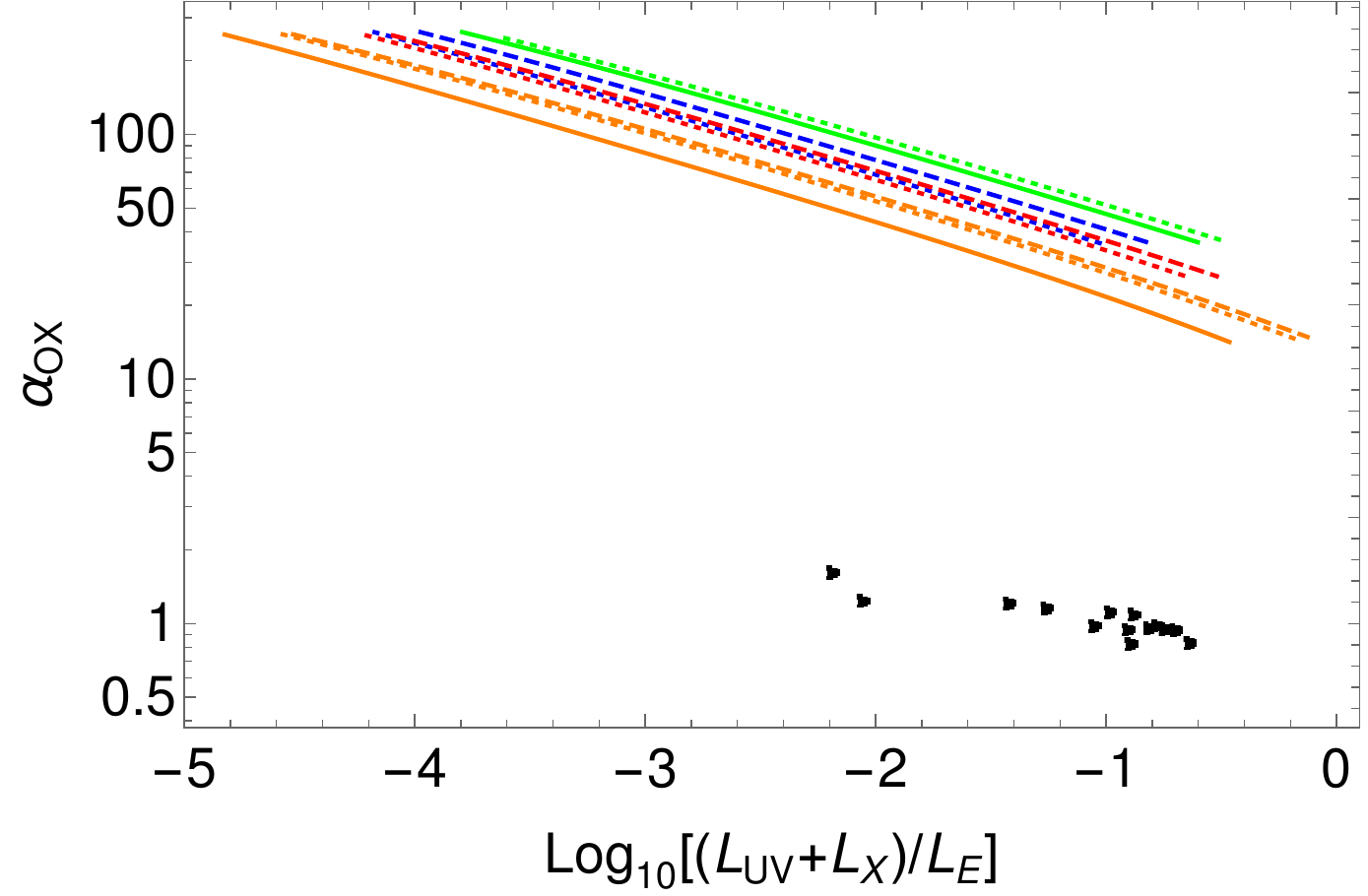}}
\subfigure[XMMSL1J0740]{\includegraphics[scale=0.55]{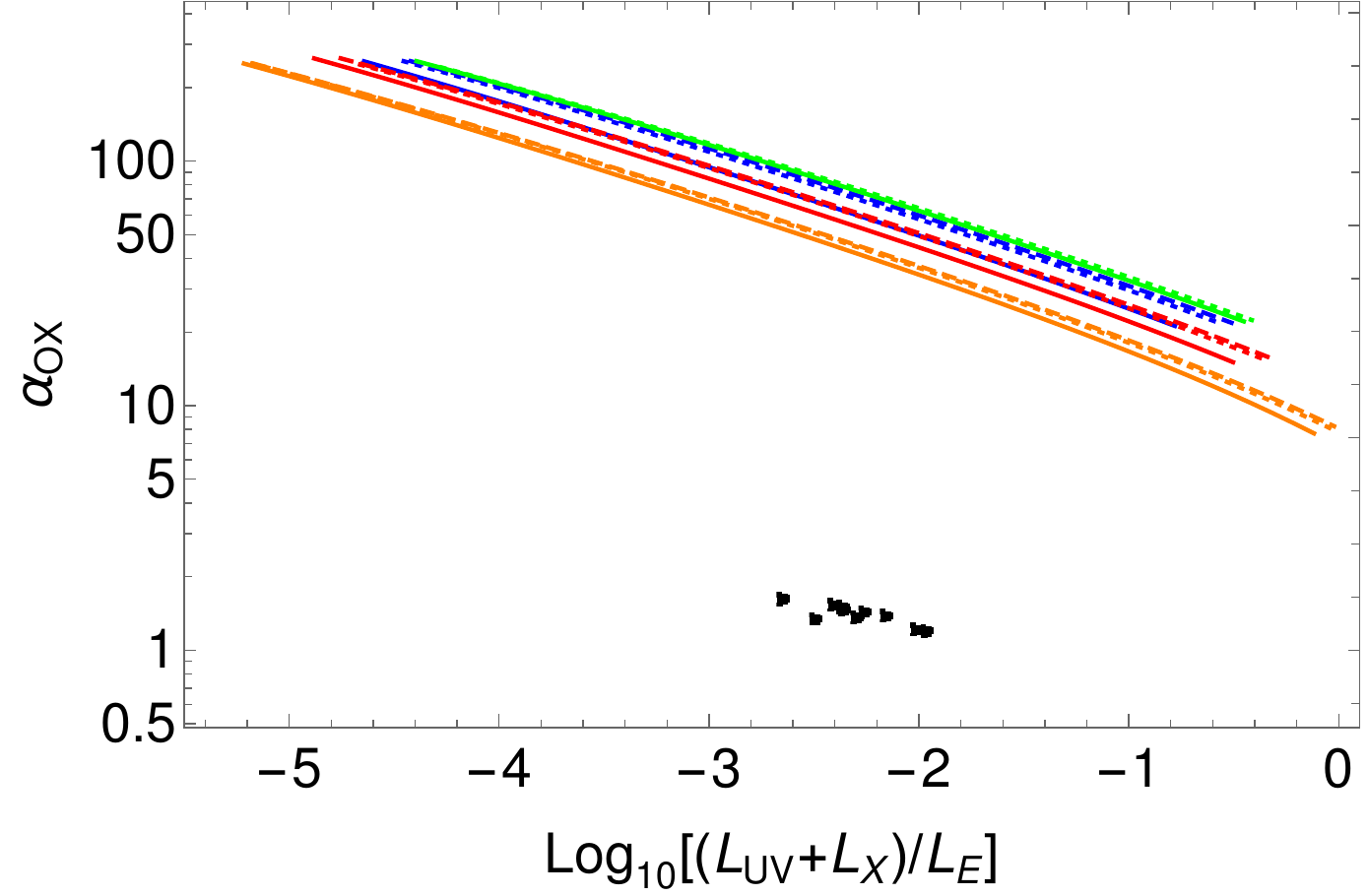}}
\end{center}
\caption{\label{ssd} The spectral index $\alpha_{\text{OX}}$ and the luminosities are calculated for the standard accretion model of \citet{1973A&A....24..337S}. In (a), the black hole mass is $M_{\bullet,6} = 10^{1.79} $ and the blue solid line corresponds to $m = 1,~j =j_m = 0.78 ,~q = 2$. In (b), the black hole mass is $M_{\bullet,6} = 10^{0.71} $ and the blue solid line corresponds to $m = 1,~j =j_m = 0 ,~q = 2$. In (c), the black hole mass is $M_{\bullet,6} = 10^{1.05} $ and the blue solid line corresponds to $m = 1,~j =j_m = 0 ,~q = 2$. The blue solid,  dotted and dashed lines correspond to $m = 1,~5,~{\text{and}}~10$ respectively calculated for their corresponding $j_m = 0.78,~0,~{\text{and}}~0$, whereas the red and orange lines corresponds to $j = 0.5~{\text{and}}~ 0.9$. The green solid and dashed lines correspond to $q = 5 ~{\text{and}}~10$ keeping other parameters the same as for the blue dashed line. The black points are the observed data taken from \citet{2020MNRAS.497L...1W}. See section \ref{modcomp} for details. }
\end{figure}

We begin with the steady accretion model for TDEs. The standard steady accretion model was constructed for an alpha viscosity dominated by gas pressure with zero torque at the inner boundary \citep{1973A&A....24..337S}. The inner radius of the disk is taken to be the innermost stable circular orbit (ISCO) given by equation (\ref{zjb}), and the outer radius is taken to be $r_{\text{out}} = q r_t$. Generally, the angular momentum conservation of the infalling debris results in $q = 2$ \citep{1999ApJ...514..180U}, but here we consider it to be an unknown free parameter. We calculate the X-ray and UV luminosities following blackbody emissions. The mass accretion rate, which is a constant parameter here, is taken in the range $\dot{M} \in \{10^{-6},~1\}\dot{M}_E$, where $\dot{M}_E$ is the Eddington mass accretion rate calculated for a radiative efficiency of $0.1$ and the parameter $q \in \{2,~10\}$. The $\alpha_{\text{OX}}$ versus luminosity is shown in FIG.~\ref{ssd}. Over the significant range of stellar mass, black hole spin, and $q$ considered, the $\alpha_{\text{OX}}$ is higher than the observed values for all the sources. 

\begin{figure}
\begin{center}
\subfigure[XMMSL2J1446]{\includegraphics[scale=0.55]{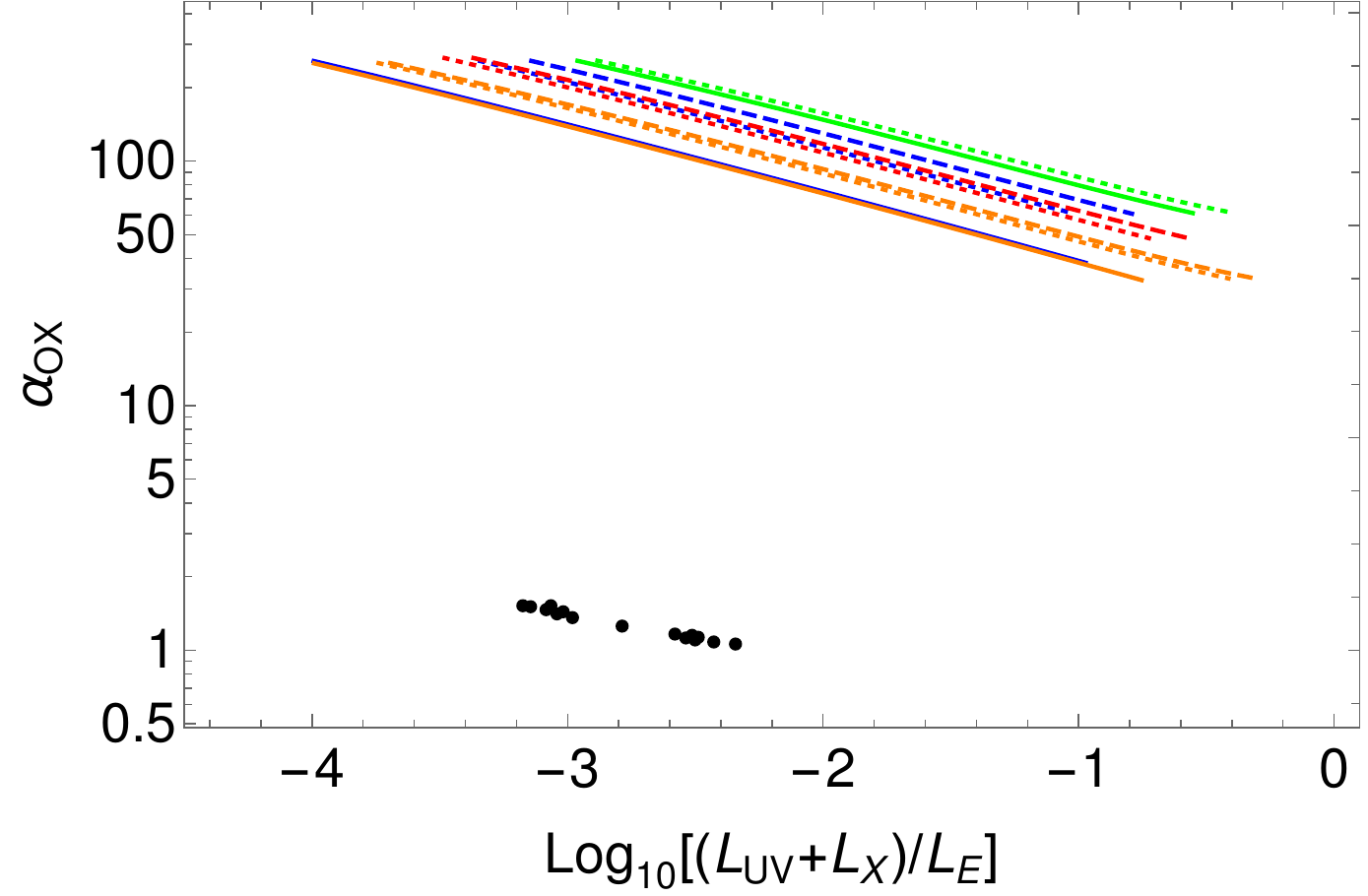}}
\subfigure[XMMSL1J1404]{\includegraphics[scale=0.55]{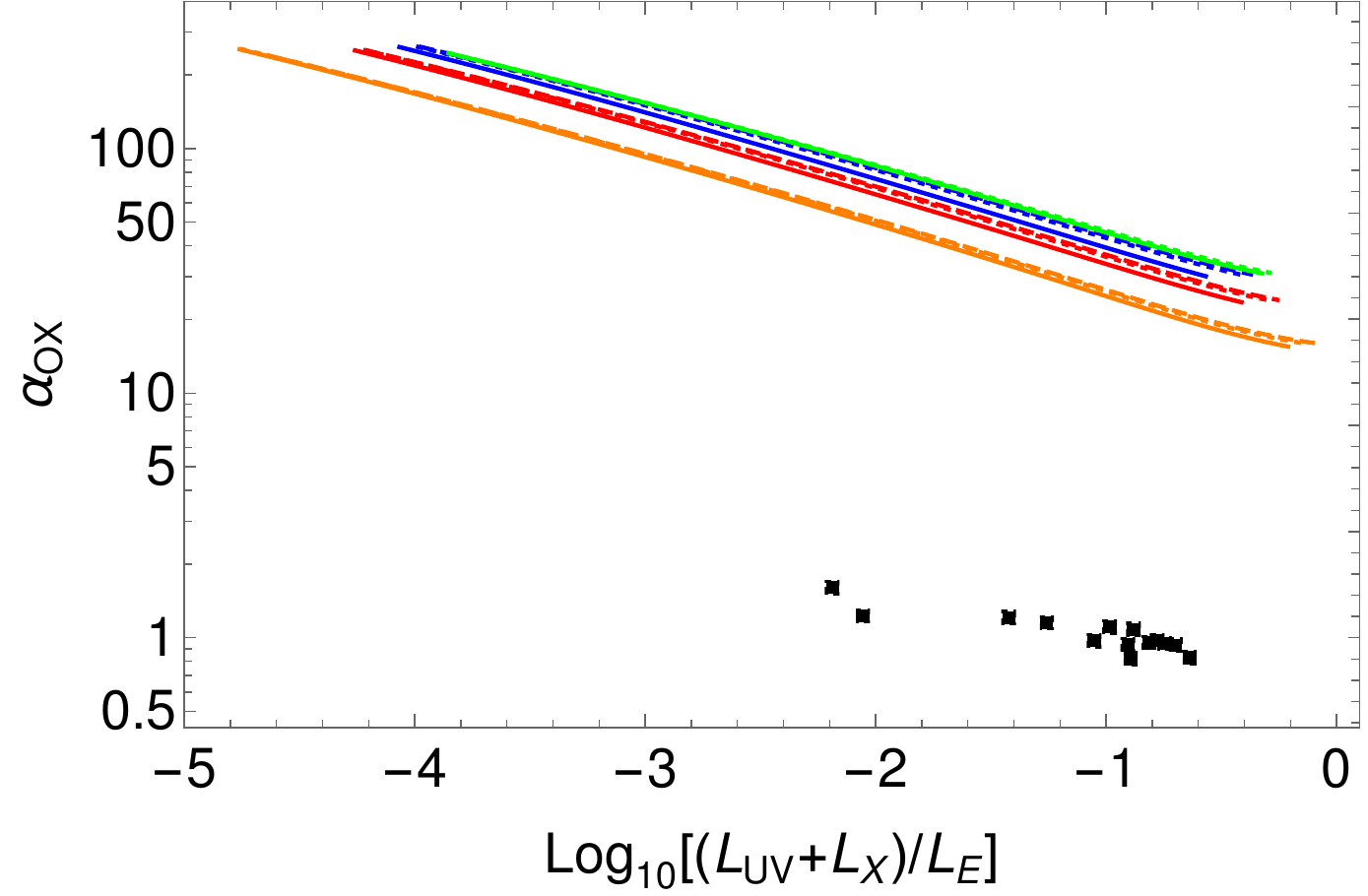}}
\subfigure[XMMSL1J0740]{\includegraphics[scale=0.55]{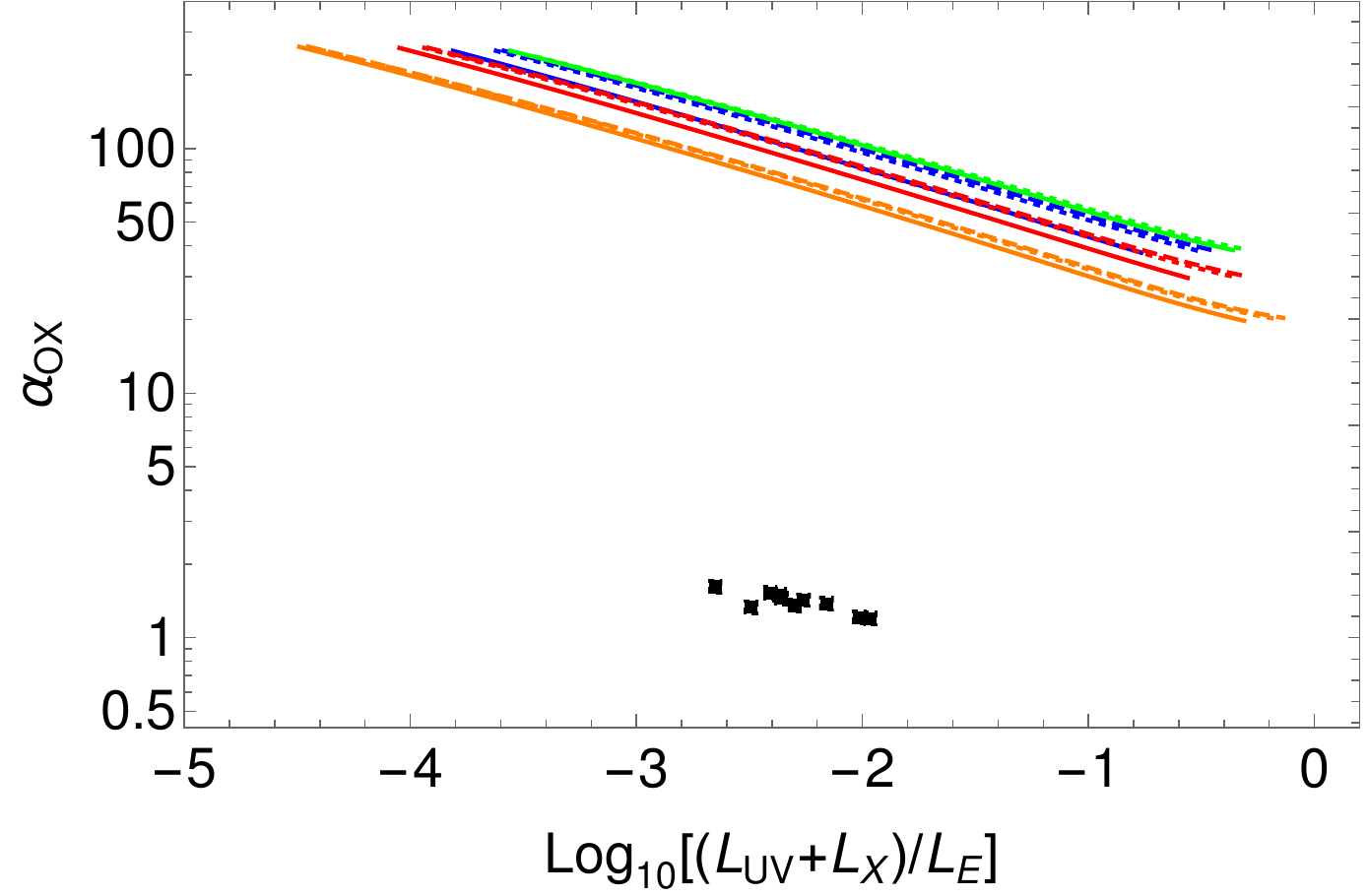}}
\end{center}
\caption{\label{sq} The spectral index $\alpha_{\text{OX}}$ and the luminosities are calculated for the steady slim accretion model of \citet{2009MNRAS.400.2070S}. In (a), the black hole mass is $M_{\bullet,6} = 10^{1.79} $ and the blue solid line corresponds to $m = 1,~j =j_m = 0.78 ,~q = 2$. In (b), the black hole mass is $M_{\bullet,6} = 10^{0.71} $ and the blue solid line corresponds to $m = 1,~j =j_m = 0 ,~q = 2$. In (c), the black hole mass is $M_{\bullet,6} = 10^{1.05} $ and the blue solid line corresponds to $m = 1,~j =j_m = 0 ,~q = 2$. The blue solid, dotted and dashed lines correspond to $m = 1,~5,~{\text{and}}~10$ respectively calculated for their corresponding $j_m = 0.78,~0,~{\text{and}}~0$, whereas the red and orange lines corresponds to $j = 0.5~{\text{and}}~ 0.9$. The green solid and dashed lines correspond to $q = 5 ~{\text{and}}~10$ keeping other parameters the same as for the blue dashed line. The black points are the observed data taken from \citet{2020MNRAS.497L...1W}. See section \ref{modcomp} for details.}
\end{figure}

Next, we consider the steady slim disk model of \citet{2009MNRAS.400.2070S} that includes the advection and alpha viscosity with radiation pressure. They obtained the temperature profile of the disk and the luminosities are calculated assuming blackbody emissions. This model is important for the accretion close to the Eddington limit where the pressure is dominated by radiation pressure. Here also, we take the inner radius of the disk to be ISCO and the outer radius to be $r_{\text{out}} = q r_t$. The mass accretion rate which is a constant parameter here is taken in the range $\dot{M} \in \{10^{-6},~1\}\dot{M}_E$, and the parameter $q \in \{2,~10\}$. The obtained values of the spectral index as shown in FIG.~\ref{sq} are higher than the observed values for the wide ranges of the parameters. This implies that the slim disk model is also inadequate to explain the observations. 

\begin{figure*}
\begin{center}
\subfigure[XMMSL2J1446]{\includegraphics[scale=0.5]{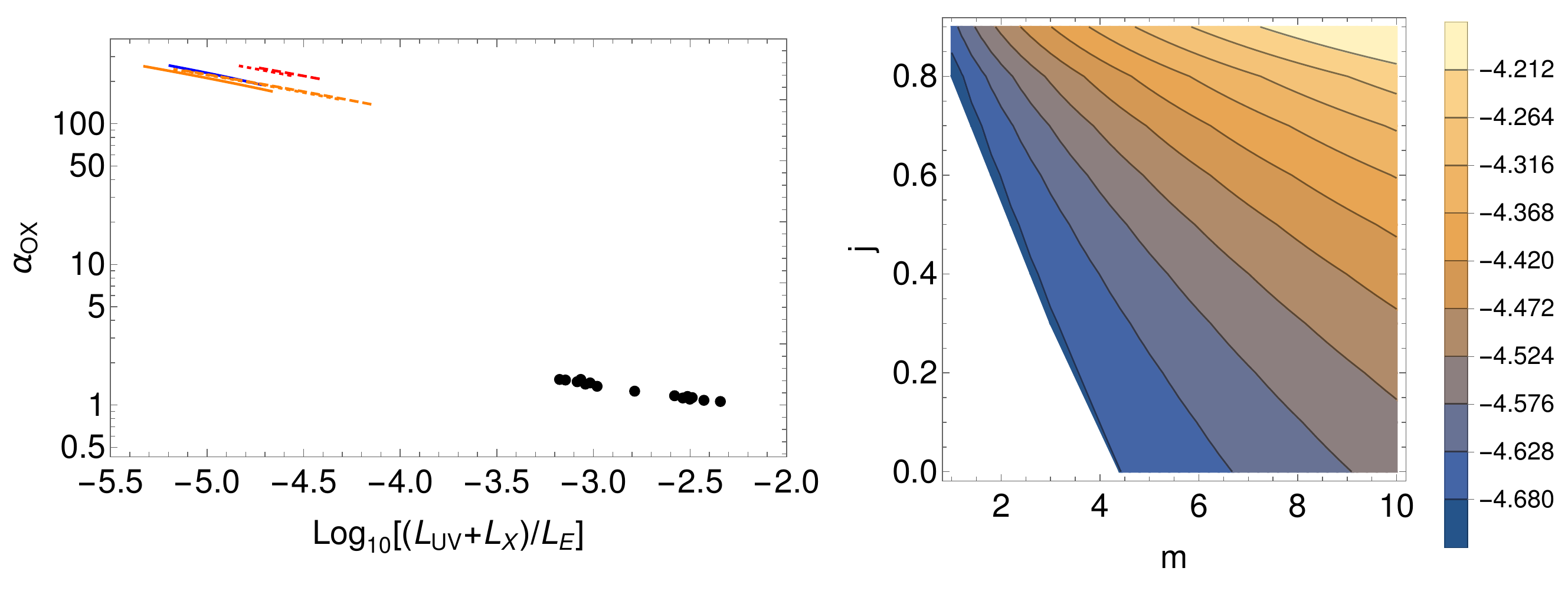}}
\subfigure[XMMSL1J1404]{\includegraphics[scale=0.5]{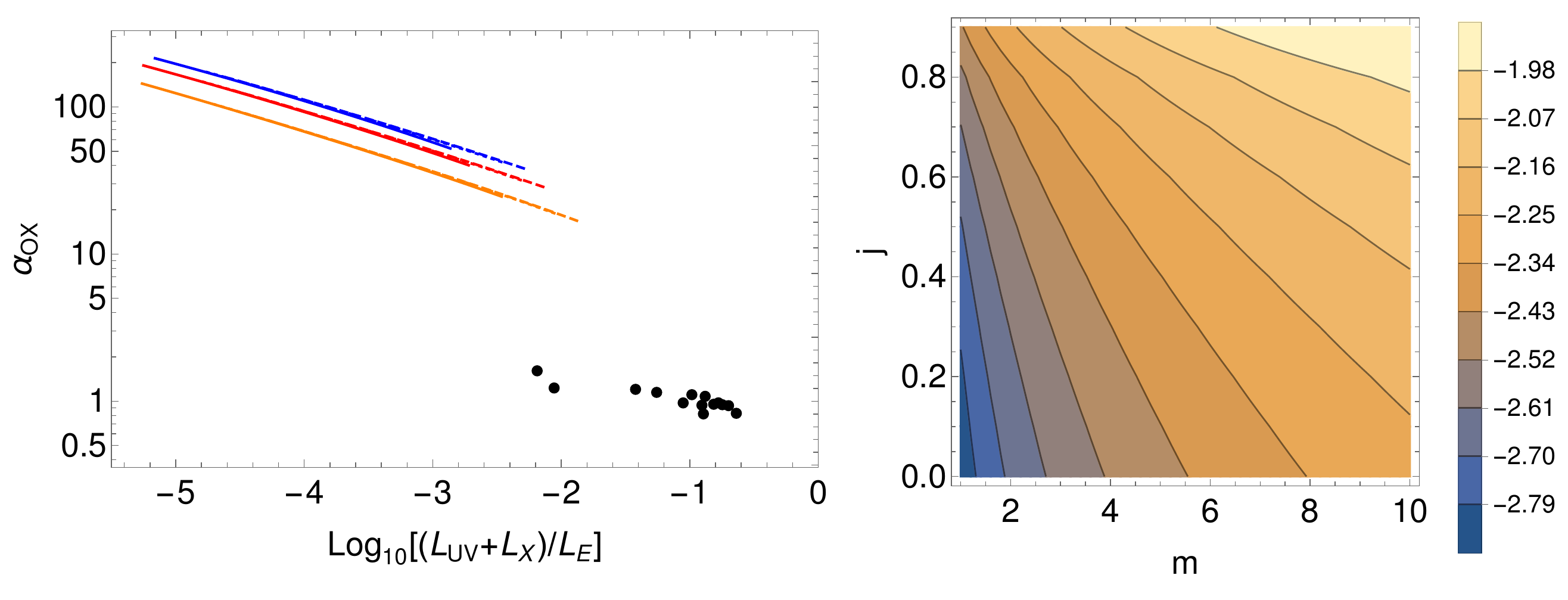}}
\subfigure[XMMSL1J0740]{\includegraphics[scale=0.5]{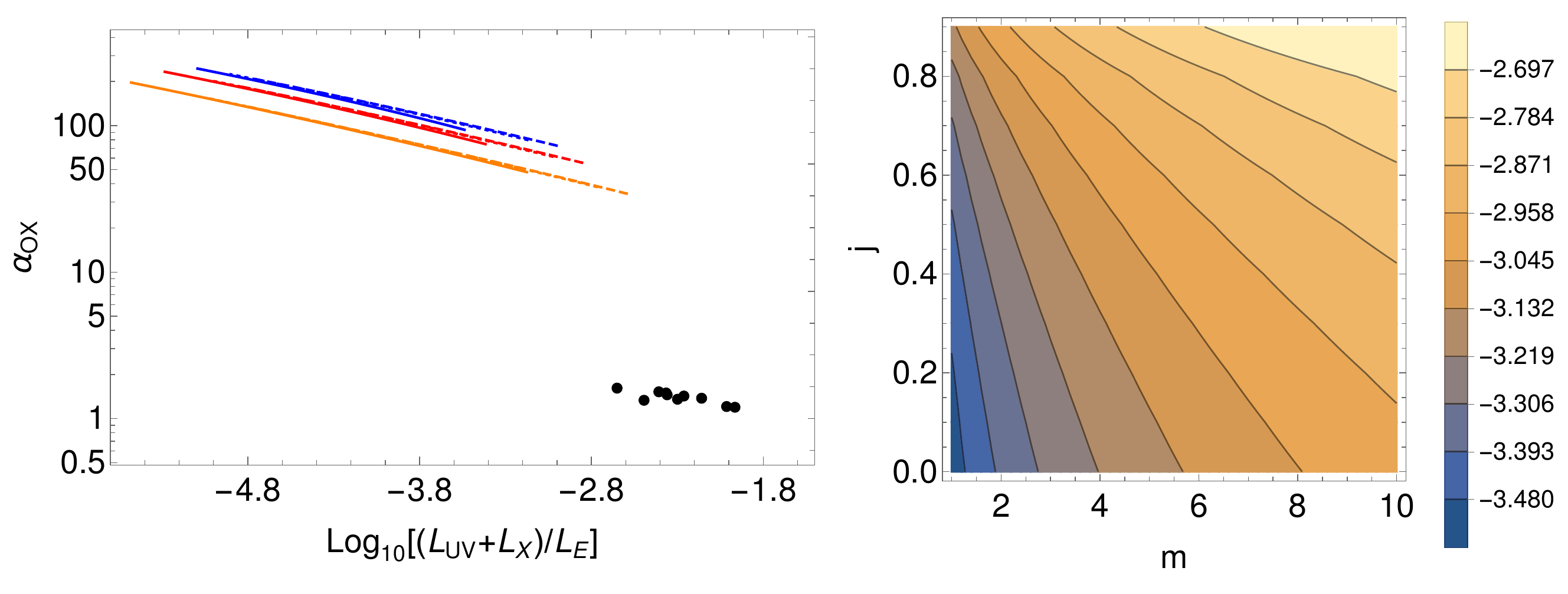}}
\end{center}
\caption{\label{can} Left: The spectral index $\alpha_{\text{OX}}$ and the luminosities are calculated for the time-dependent self-similar accretion model of \citet{1990ApJ...351...38C}. In (a), the black hole mass is $M_{\bullet,6} = 10^{1.79} $ and the blue solid line corresponds to $m = 1,~{\text{and}}~j =j_m = 0.78$. In (b), the black hole mass is $M_{\bullet,6} = 10^{0.71}$ and the blue solid line corresponds to $m = 1,~{\text{and}}~j =j_m = 0$. In (c), the black hole mass is $M_{\bullet,6} = 10^{1.05} $ and the blue solid line corresponds to $m = 1,~{\text{and}}~j =j_m = 0$. The blue solid,  dotted and dashed lines correspond to $m = 1,~5,~{\text{and}}~10$ respectively calculated for their corresponding $j_m = 0.78,~0,~{\text{and}}~0$, whereas the red and orange lines correspond to $j = 0.5~{\text{and}}~ 0.9$. Some lines are missing in (a) because the 2 keV X-ray luminosity is too small that the spectral index is very high. The black points are the observed data taken from \citet{2020MNRAS.497L...1W}. Right: The maximum value of Eddington luminosity ratio $(L_{UV} + L_X )/L_E$ from the model as a function of black hole spin and stellar mass for all the three sources.  See section \ref{modcomp} for details.}
\end{figure*}

The sub-Eddington steady accretion models are unable to explain the observed spectral index. We now look for the time-dependent accretion models in the sub-Eddington phase with and without a fallback. The alpha-viscous stress when the pressure is dominated by gas pressure is given by $\Pi_{r\phi} \propto \Sigma^{5/3} r^{-1/2}$ \citep{1973A&A....24..337S}. With power-law viscous stress, \citet{1990ApJ...351...38C} constructed a self-similar solution for a disk with total angular momentum constant and without fallback, where the disk outer radius increases with time as $r_{\text{out}} \propto t^{3/8}$ where $t$ is the time parameter. We use their self-similar solution and assume that the total mass of the initial disk is $M_{\star}/2$ and calculate the initial outer radius by assuming that the total angular momentum of the disk is equal to the total angular momentum of the bound disrupted debris given by $\sqrt{2 G M_{\bullet} r_t} M_{\star}/2$. Thus, the unknown parameters for this model are $m$ and $j$. We calculate the spectral index and the luminosities by varying the time $t$ over a long duration during which both surface density and outer radius evolve. FIG.~\ref{can} shows that the spectral index is higher and the luminosities are smaller than the observed values. The maximum value of luminosity ratio $(L_{UV} + L_X )/L_E$ from the model is less than the observed value and thus the spectral index curve does not cover the entire observed range for all the parameters.

\begin{figure*}
\begin{center}
\subfigure[XMMSL2J1446]{\includegraphics[scale=0.5]{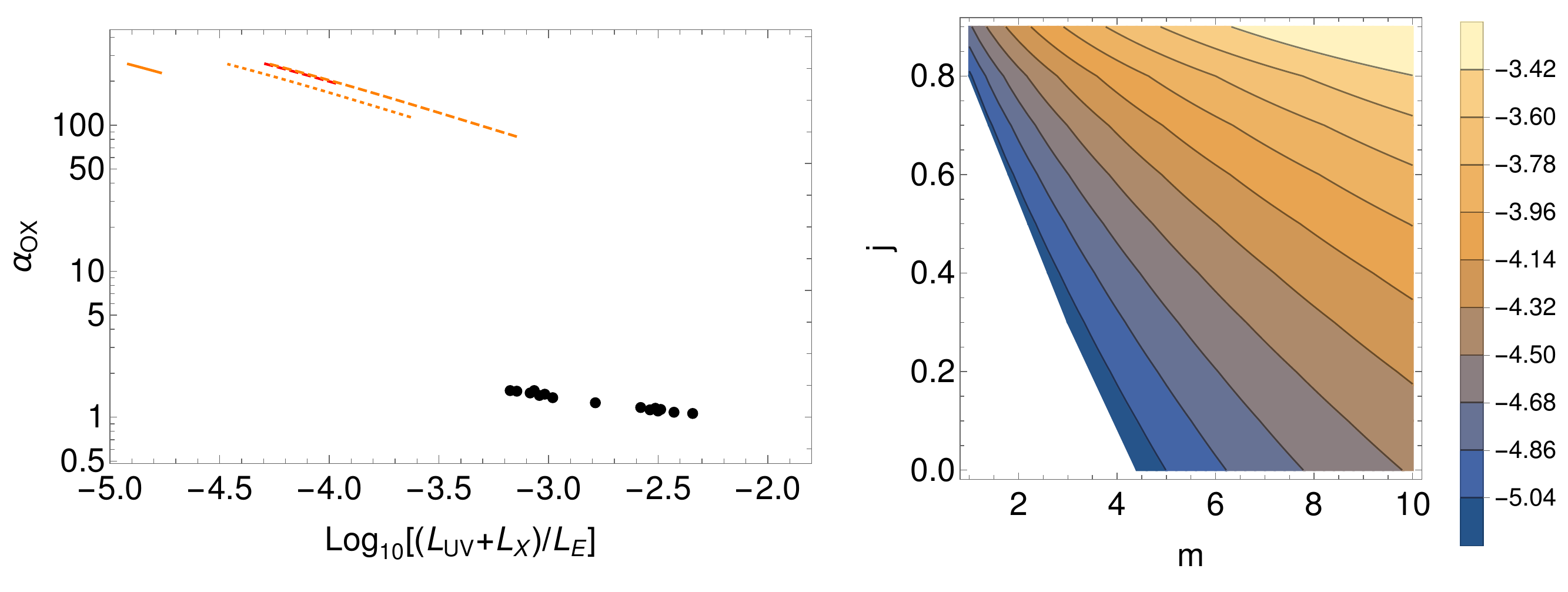}}
\subfigure[XMMSL1J1404]{\includegraphics[scale=0.5]{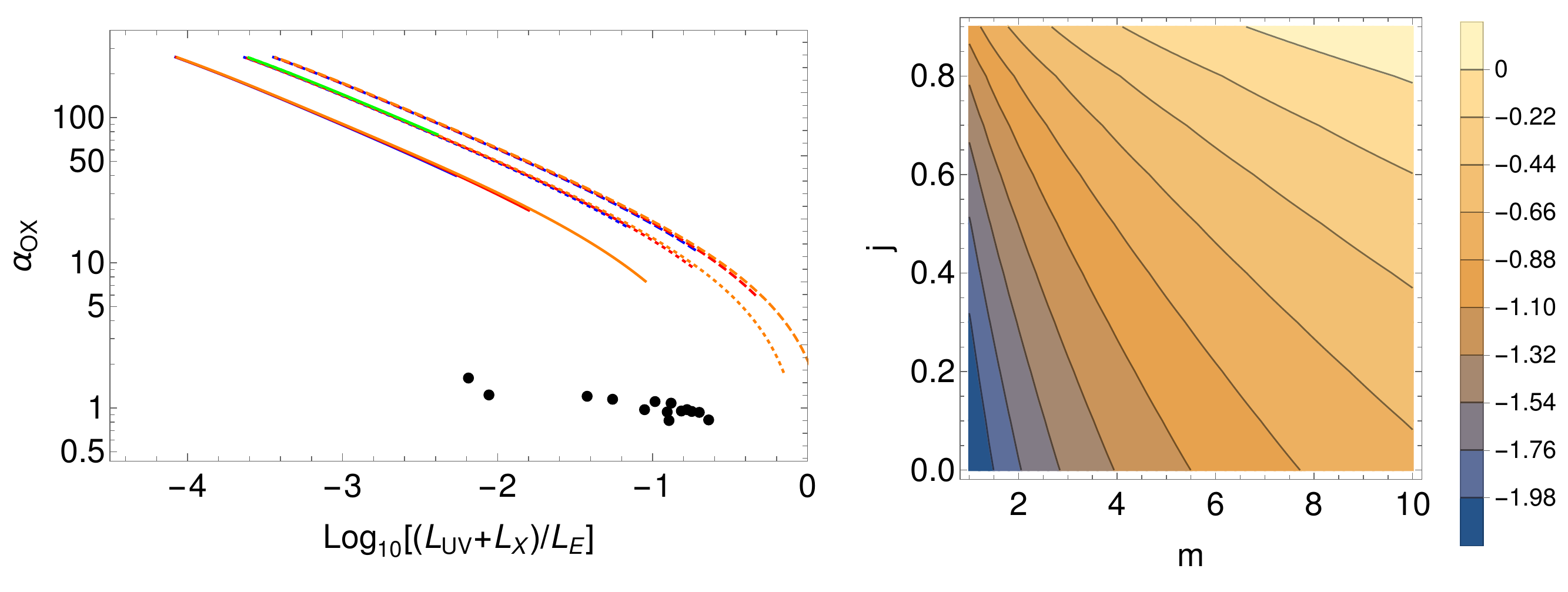}}
\subfigure[XMMSL1J0740]{\includegraphics[scale=0.5]{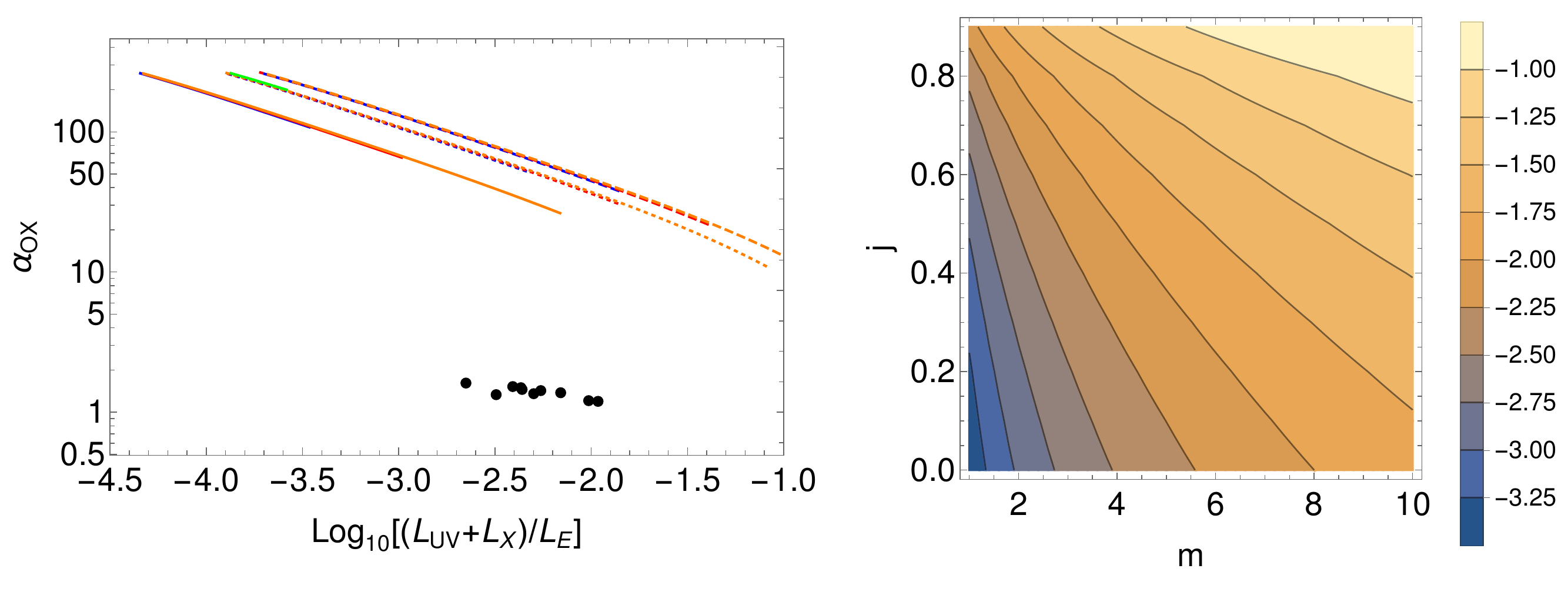}}
\end{center}
\caption{\label{MM} Left: The spectral index $\alpha_{\text{OX}}$ and the luminosities are calculated for the time-dependent self-similar accretion model with fallback of \citet{2021NewA...8301491M}. In (a), the black hole mass is $M_{\bullet,6} = 10^{1.79} $ and the blue solid line that corresponds to $m = 1,~j =j_m = 0.78 ,~q = 2$ has very large values of $\alpha_{\text{OX}}$ due to highly small X-ray luminosity and thus not visible here. In (b), the black hole mass is $M_{\bullet,6} = 10^{0.71} $ and the blue solid line corresponds to $m = 1,~j =j_m = 0 ,~q = 2$. In (c), the black hole mass is $M_{\bullet,6} = 10^{1.05} $ and the blue solid line corresponds to $m = 1,~j =j_m = 0 ,~q = 2$. The blue solid, dotted and dashed lines correspond to $m = 1,~5,~{\text{and}}~10$ respectively calculated for their corresponding $j_m = 0.78,~0,~{\text{and}}~0$, whereas the red and orange lines correspond to $j = 0.5~{\text{and}}~ 0.9$. The green solid and dashed lines correspond to $q = 5 ~{\text{and}}~10$ keeping other parameters the same as for the blue dashed line. Some lines are very closeby like blue, red, and orange lines in (b). The black points are the observed data taken from \citet{2020MNRAS.497L...1W}. Right: The maximum value of Eddington luminosity ratio $(L_{UV} + L_X )/L_E$ from the model as a function of black hole spin and stellar mass for all the three sources. See section \ref{modcomp} for details.}
\end{figure*}

We next consider the time-dependent accretion model with fallback constructed for a sub-Eddington disk with an alpha viscosity. \citet{2021NewA...8301491M} developed a power-law self-similar solution for the same viscous stress used by \citet{1990ApJ...351...38C}. They assumed that the infalling debris forms an initial seed disk in an initial time (calculated self-consistently in the formulation) whose mass is equal to the debris mass that has infall by that time. The disk then evolves in the effect of viscous accretion onto the black hole and the later debris infalls. Then, by considering that the rate of disk mass change ($\dot{M}_d$; where $M_d$ is disk mass) is equal to the mass fallback rate ($\dot{M}_{\text{fb}}$) minus the mass accretion rate ($\dot{M}_a$), they estimated the evolution of the outer radius. The initial outer radius of the disk is taken to be $r_{\text{out}} = q r_{\text{in}}$, where $q$ is a free parameter and inner radius $r_{\text{in}}$ is taken to be an ISCO radius. To calculate the mass fallback rate, we consider the stellar orbit before disruption to be parabolic and the specific energy of the disrupted debris is governed by the variation of the black hole potential across the star and the tidal spin-up of the star as a result of the tidal interaction (\citealp{2020MNRAS.496.1784M}: model MFR1). The disrupted debris is assumed to return to the pericenter following a Keplerian orbit and the mass fallback rate is calculated following the semi-analytic formulation given in \citet{2009MNRAS.392..332L}. Then, the luminosities are calculated assuming a blackbody emission. With the given black hole mass, the unknown parameters here are $m$, $j$, and $q$ whose ranges are the same as those considered for other models. We calculate the spectral index and the luminosities by varying the time $t$ over a long duration during which both surface density and outer radius evolve. FIG.~\ref{MM} shows that there is a discrepancy between the observed values and the model values considered over a significant range of $m$, $j$, and $q$. The maximum value of luminosity ratio $(L_{UV} + L_X )/L_E$ from the model is less than the observed value and thus the spectral index curve does not cover the entire observed luminosity range for all the parameters. 

\begin{figure*}
\begin{center}
\includegraphics[scale=0.75]{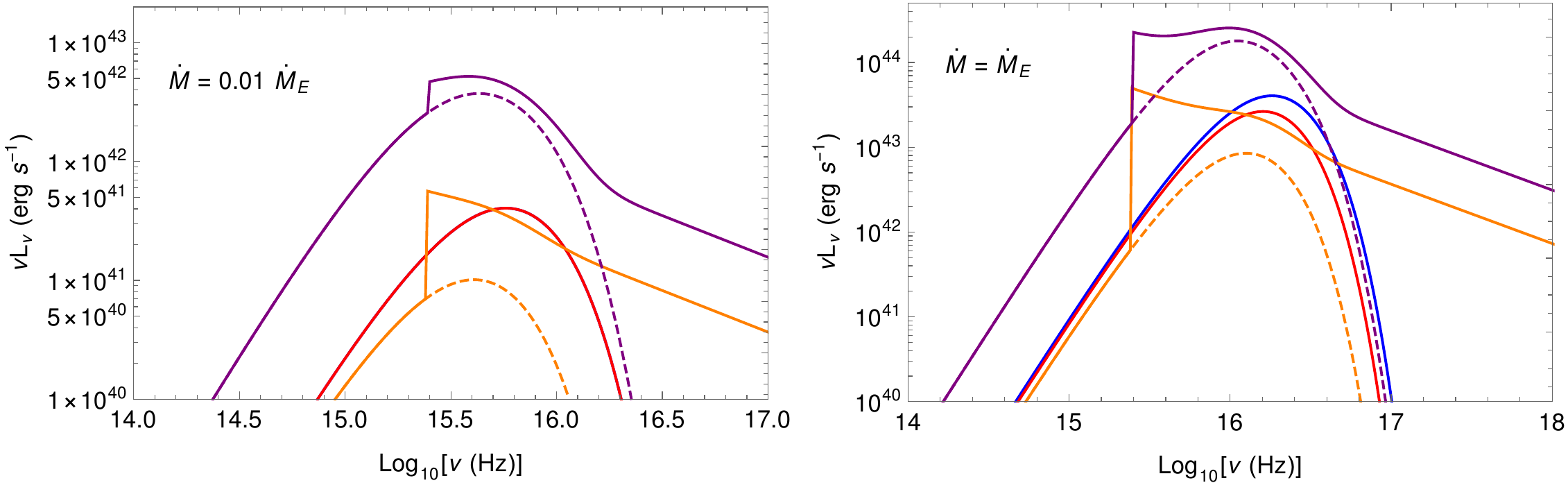}
\end{center}
\caption{\label{J1404comp} We compare the steady relativistic (purple) and non-relativistic (orange) disk-corona models with the standard steady accretion model (blue) of \citep{1973A&A....24..337S} and steady advective slim disk model (red) of \citet{2009MNRAS.400.2070S} for XMMSL1J1404. The X-ray luminosity from the corona is taken in range 0.01 - 10 keV and induces a sudden change in spectral luminosity. The dashed lines represent the disk blackbody emissions for the disk-corona models. The black hole mass $M_{\bullet} = 10^{6.71} M_{\odot}$, $j = 0$, $M_{\star} = M_{\odot}$, disk inner radius $r_{\rm in} = r_{\rm ISCO}$ and outer radius $r_{\rm out} = 2 r_t$. For the disk-corona cases, $\alpha_0 = 0.5$, $\eta = 0.5$ and $a_d = 0.5$. The blue and red lines overlap for $\dot{M} = 0.01 \dot{M}_E$ as the slim disk model is equivalent to thin disk for low mass accretion rates due to insignificant advection energy flux. See section \ref{modcomp} for details.}
\end{figure*}

We have seen that the steady and time-dependent accretion models are insufficient to match the observed $\alpha_{\text{OX}}$. In section \ref{rdcm}, we have modeled a relativistic disk-corona model, and a non-relativistic formulation is shown in appendix \ref{ndcm}. Since the disk-corona models are steady structured, we compare models with the standard steady accretion model of \citep{1973A&A....24..337S} and steady advective slim disk model of \citet{2009MNRAS.400.2070S} for two mass accretion rates. The time-dependent accretion models have varying mass accretion rates in time and radius and are not compared with the steady accretion models. However, we have shown that they are unable to explain the observed spectral index. The relativistic disk-corona model shows a higher spectral luminosity compared to the non-relativistic models. The relativistic effects dominate near ISCO where the effective temperature peaks and shows a significant impact on the spectral luminosity. The energy transfer to the corona reduces the radiative energy flux causing a decrease in disk luminosity as can be seen from the blue, red, and orange lines corresponding to the non-relativistic models. The X-ray luminosity from the corona is taken in the range of 0.01 - 10 keV. The total luminosity shown by solid lines shows an increment in the luminosity due to emission from the corona. The luminosity from the disk is lower than that from the corona in X-ray bands. The emission from the corona increases the 2 keV X-ray luminosity resulting in a reduction in the spectral index. The increment in the luminosity due to the corona is not smooth and that is because we have neglected the details of the physical cooling mechanism in the corona and approximated the coronal emission by a simple power-law X-ray emission. The energy transferred to the corona heats the medium and the corona cools via bremsstrahlung, synchrotron, and Compton processes. The detailed modeling of accretion disk-corona with these processes will be done in the future.

We now fit the relativistic disk model with corona discussed in \ref{rdcm}. With the given black hole mass and photon index, we have seven unknown parameters given by $\alpha_0$, $\mu$, $\eta$, $a_d$, $m$, $j$ and $q = r_{\text{out}}/r_{t}$. We fit the models to the data using the $\chi^2$ minimization given by \citep{Grossman1971}

\begin{multline}
\chi^2 = \sum_{i = 1}^{i = N} \frac{(\alpha_{\text{OX,M}}^{i}(\dot{M})-\alpha_{\text{OX,O}}^{i})^2}{\sigma_{\text{OX,i}}^2} +  \\ \frac{(l_{\text{UV+X,M}}^{i}(\dot{M})-l_{\text{UV+X,O}}^{i})^2}{\sigma_{\text{L,i}}^2},
\label{chi}
\end{multline}

\noindent where $N$ is the number of data points, $l_{\text{UV+X}} = \log_{10}[(L_X+L_{\text{UV}})/L_E]$, the subscript $M$ and $O$ signifies the model and the observed values, and $\sigma_{\text{OX,i}}$ and $\sigma_{\text{L,i}}$ is the observational errors. The observational errors for a single data point for each source are given in \citet{2020MNRAS.497L...1W}, and we assume each data point for a given source to have the same error. The $\{\sigma_{\text{OX}},~\sigma_{\text{L}}\}$ is $\{ 0.122,~0.503\}$ (XMMSL1J0740), $\{0.09,~0.69\}$ (XMMSL2J1446), and $\{0.115,~0.472\}$ (XMMSL1J1404). Since the $\alpha_{\text{OX}}$ and the luminosity $l_{\text{UV+X}}$ are function of $\dot{M}$, we perform $\chi^2$ minimization for individual points by varying the mass accretion rate in the range given by $\dot{M} \in \{10^{-5},~10^2\} \dot{M}_E$, where $\dot{M}_E$ is the mass Eddington rate calculated for efficiency $0.1$. Then we sum the $\chi^2$ from individual points to get the total $\chi^2$. This process is followed until we get the minimum of total $\chi^2$. We use the numerical optimization package of Wolfram Mathematica\footnote{\url{https://reference.wolfram.com/language/tutorial/ConstrainedOptimizationGlobalNumerical.html}} to find the solutions. 

\begin{figure}
\begin{center}
\includegraphics[scale=0.8]{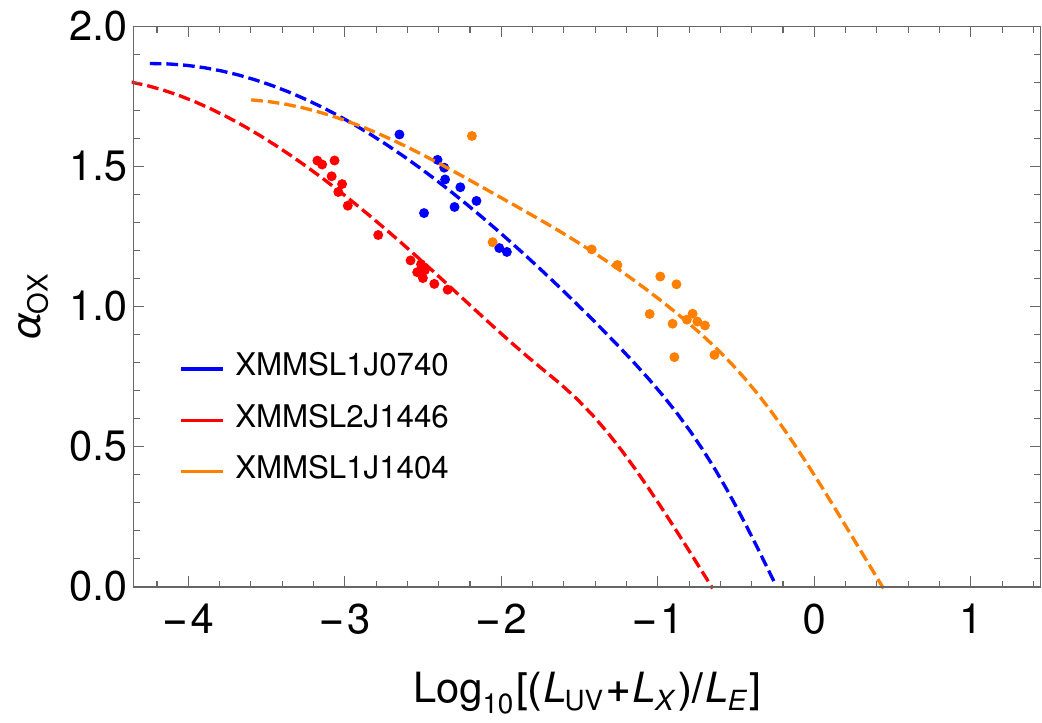}
\end{center}
\caption{\label{fitplt} The figure shows the relativistic model fit to the observations of the considered TDE sources. The obtained values are shown in Table \ref{rres} for the relativistic model. }
\end{figure}

\begin{figure*}
\begin{center}
\includegraphics[scale=0.27]{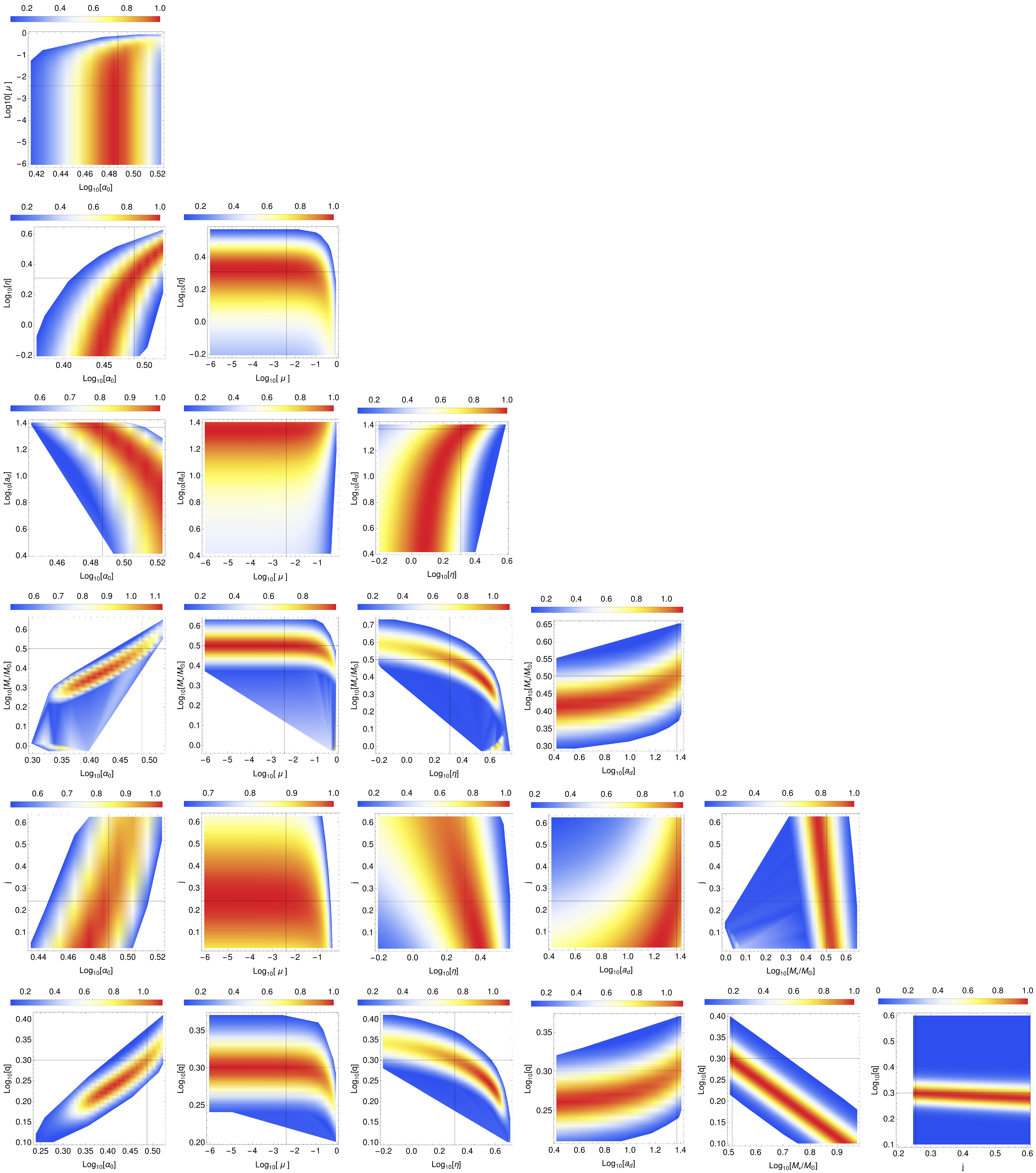}
\end{center}
\caption{\label{XM1446L} The density plots show the relative likelihood $\bar{\mathcal{L}} = \mathcal{L} / \mathcal{L}_p > 0.1$ for XMMSL2J1446, where $\mathcal{L}_p$ is the likelihood at the obtained parameters values which are shown by solid black lines. The bar legend above individual plots shows the relative likelihood. See section \ref{modcomp} for details.}
\end{figure*}

\begin{figure*}
\begin{center}
\includegraphics[scale=0.27]{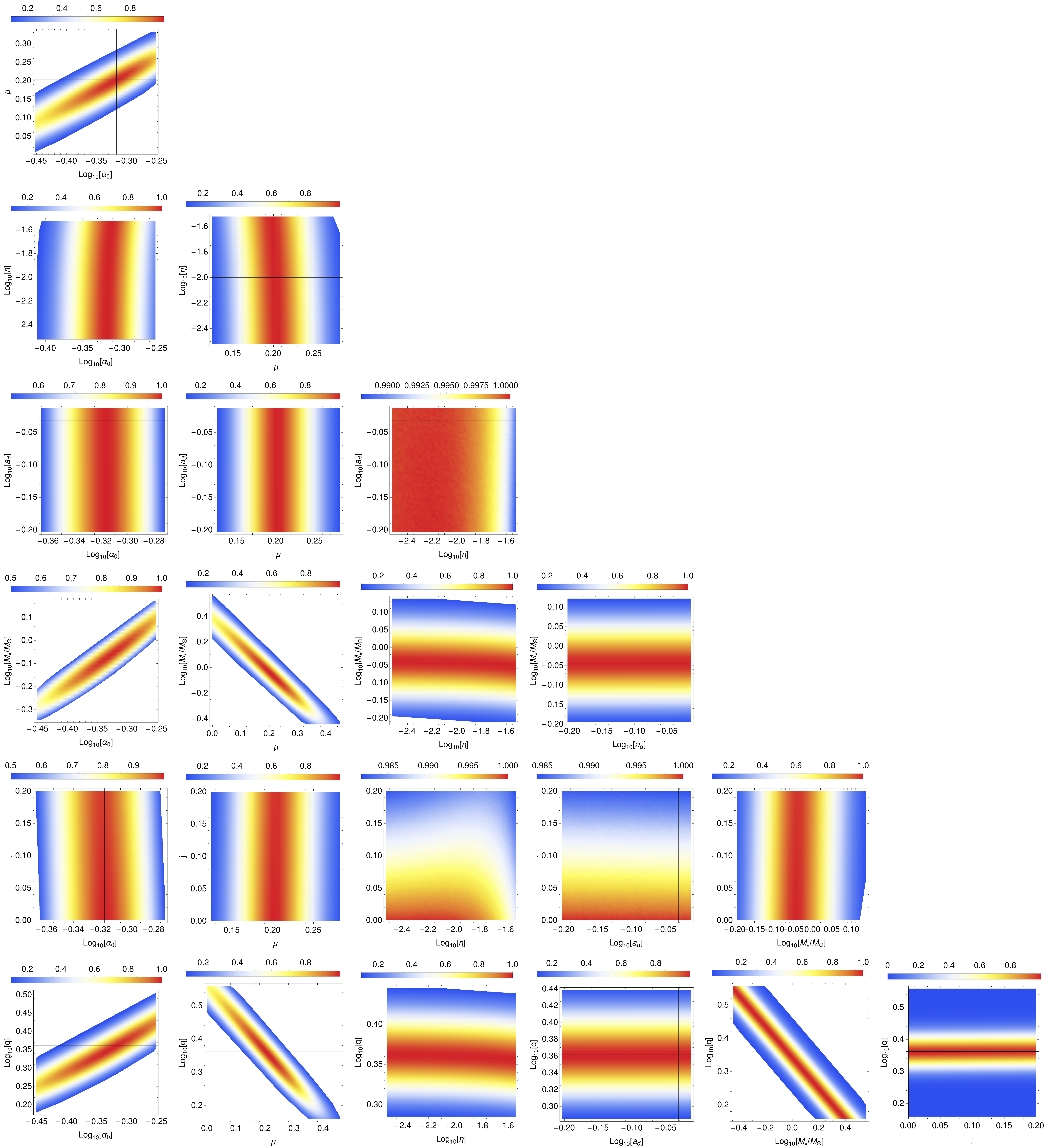}
\end{center}
\caption{\label{XM1404L} The density plots show the relative likelihood $\bar{\mathcal{L}} = \mathcal{L} / \mathcal{L}_p > 0.1$ for XMMSL1J1404, where $\mathcal{L}_p$ is the likelihood at the obtained parameters values which are shown by solid black lines. The bar legend above individual plots shows the relative likelihood. See section \ref{modcomp} for details.}
\end{figure*}

\begin{figure*}
\begin{center}
\includegraphics[scale=0.27]{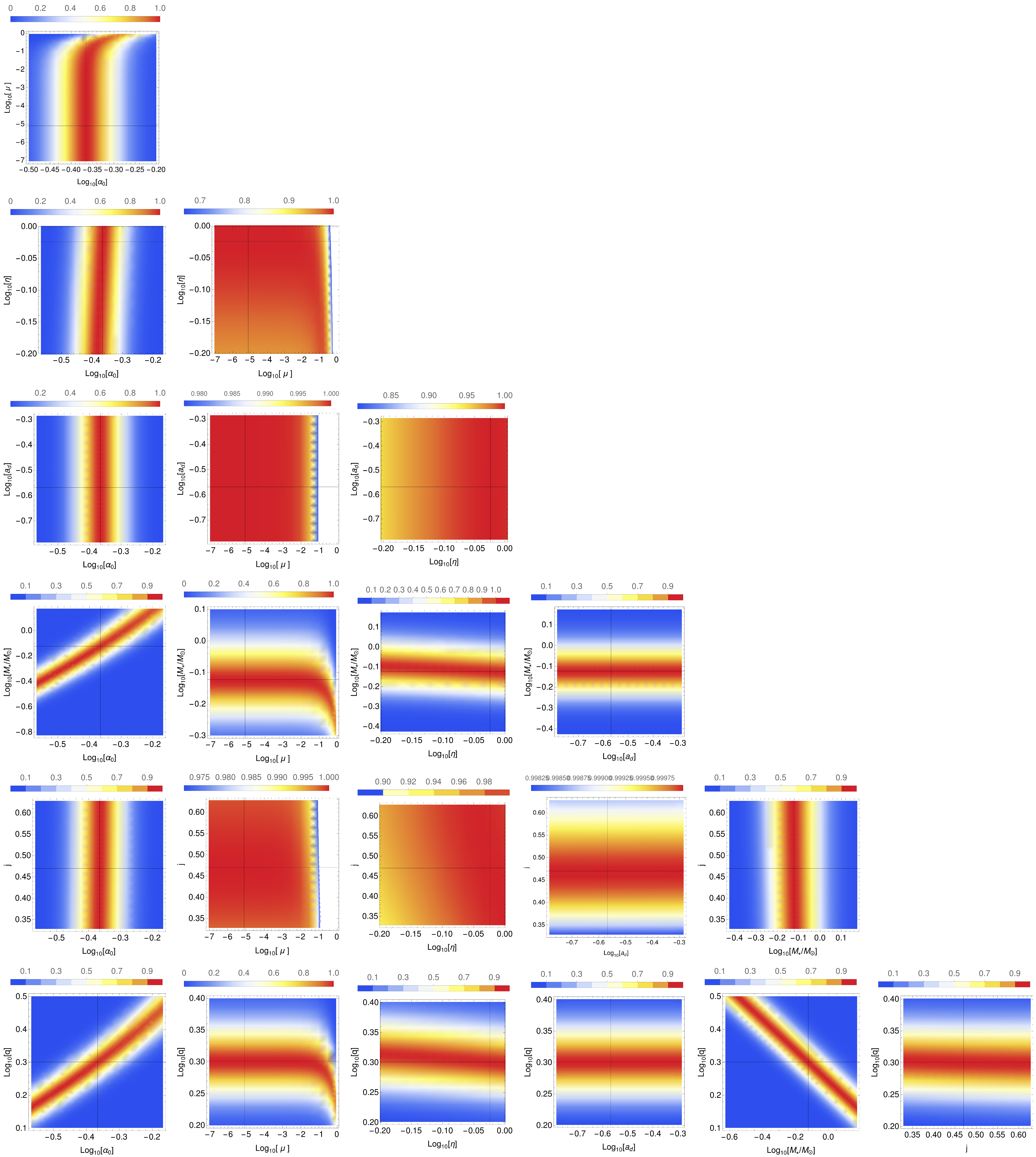}
\end{center}
\caption{\label{XM0740L} The density plots show the relative likelihood $\bar{\mathcal{L}} = \mathcal{L} / \mathcal{L}_p > 0.1$ for XMMSL1J0740, where $\mathcal{L}_p$ is the likelihood at the obtained parameters values which are shown by solid black lines. The bar legend above individual plots shows the relative likelihood. See section \ref{modcomp} for details.}
\end{figure*}

FIG.~\ref{fitplt} shows the relativistic model fit to the observations for the three TDE sources and the obtained parameters are given in Table \ref{rres}. The relative likelihood distribution (likelihood normalized to the likelihood for the $\chi^2$ given in Table \ref{rres}) for XMMSL2J1446, XMMSL1J1404, and XMMSL1J0740 are shown in FIGs.~ \ref{XM1446L}, \ref{XM1404L} and \ref{XM0740L} respectively. The obtained parameters are within 90\% of the relative likelihood and suggest that they are good estimates. However, the relative likelihood is not well constrained along the downward component, $\eta$, and disk albedo, $a_d$, resulting in significant error estimations. The obtained value of $\mu$ is well constrained and is small for all the cases which implies that the viscous stress is dominated by the total pressure. We use the obtained mean solution to estimate the luminosity and mass accretion rate. 

\begin{table*}
\caption{\label{rres} The Parameter values obtained by minimizing the equation (\ref{chi}) using the relativistic disk-corona model along with the reduced chi square (see section \ref{rdcm}) for the considered TDE sources.}
\begin{ruledtabular}
\tiny
\begin{tabular}{cccccccccc}
&&&&&&&&&\\
Sources & $\log(\alpha_0)$ & $\mu$ & $\log(\eta)$ & $\log(a_d)$ & $\log[M_{\star}/M_{\odot}]$ & $j$ & $\log(q)$ & $\chi_r^2$  \\
&&&&&&&&&\\
\hline
&&&&&&&&&\\
XMMSL1J0740 & $-0.365 \pm 0.254$ & $(7.96 \pm 2.25) \times 10^{-6}$ & $-0.0241 \pm 0.91$ & $-0.568 \pm 1.2$  & $-0.0705 \pm 0.72$  & $0.47 \pm 0.58$ & $0.30 \pm 0.04$ & 0.78  \\
&&&&&&&&&\\
\hline
&&&&&&&&&\\
XMMSL2J1446 & $-0.301 \pm 0.079$ & $10^{-4.361 \pm 1.99}$ & $-0.485 \pm 0.571$ & $-0.05 \pm 0.358$ & $0.52 \pm 0.407$ & $0.214 \pm 0.406$ & $ 0.301 \pm 0.382$ & 1.23 \\
&&&&&&&&&\\
\hline
&&&&&&&&&\\
XMMSL1J1404 & $-0.317 \pm 0.106$ & $0.203 \pm 0.06$ & $ -2 \pm 0.60$ & $ -0.031 \pm 0.81$  & $-0.041 \pm 0.15 $ & $(2.7 \pm 0.418) \times 10^{-4}$ & $0.361 \pm 0.427$ & 1.67 \\
&&&&&&&&&\\
\end{tabular}
\end{ruledtabular}
\end{table*}

\section{\label{discuss} discussion}

The disruption of stars by black holes gravitational tidal force and their accretion onto the black holes is a promising phenomenon to study the accretion dynamics through the sub and super-Eddington phases around supermassive black holes on a timescale of a few years. The circularization process of the disrupted debris to form an accretion disk is complex and still unclear as the numerical simulation studies have been limited to a few parameter ranges of black hole mass and spin, and stellar mass \citep{2015ApJ...804...85S,2016MNRAS.461.3760H}. A global study on the circularization dynamics is yet to be done. However, the performed simulations have shown that the debris could form an accretion disk and the nature of the formed disk depends on the stream-stream interactions resulting in angular momentum exchange, viscous dynamics within the debris, the thermal radiative efficiency of the debris, and on the pericenter and the eccentricity of the initial stellar orbit \citep{2020A&A...642A.111C}. We assume that the debris forms an accretion disk with an inner radius at ISCO. 

When the tidal radius lies above the ISCO, the stream-stream interactions result in an exchange of angular momentum, and the matter moves inward to form a circular disk with an inner radius of ISCO. However, when the tidal radius lies within ISCO, the infalling debris may form an accretion disk with ISCO inner radius but some fraction of the infalling debris will be plunged to the black hole due to angular momentum exchange via stream interactions. The structure of such a disk and the dynamics of circularization is uncertain and needs detailed relativistic stream modeling. Due to uncertainty in the disk formation, here, we consider only those cases where the tidal radius lies outside ISCO. This results in a constraint on black hole spin $j$ given by $j \geq j_m$, where $j_m$ is shown in FIG.~\ref{jmint}. 

We have considered the various steady and time-dependent sub-Eddington models in section \ref{modcomp}. The modeled $\alpha_{\text{OX}}$ is significantly higher than the observed values and implies that the X-ray luminosity is much smaller than the UV luminosity. If we compare the steady standard accretion model (FIG.~\ref{ssd}) and the steady slim disk model (FIG.~\ref{sq}), the $\alpha_{\text{OX}}$ obtained from the steady slim disk model is higher than that of the standard accretion model. This is because the slim disk model includes the advection which decreases the effective temperature and thus the X-ray luminosity will be smaller than that of the standard accretion model. The bolometric luminosity of a time-dependent disk without fallback is $L_b \propto t^{-1.2}$, whereas, for a disk with fallback, it is $L_b \propto t^{-1.42}$. However, the total luminosity (X-ray + UV) is smaller for a time-dependent accretion model without fallback (FIG.~\ref{can}) than with fallback (FIG.~\ref{MM}). 

\begin{figure}
\begin{center}
\includegraphics[scale=0.65]{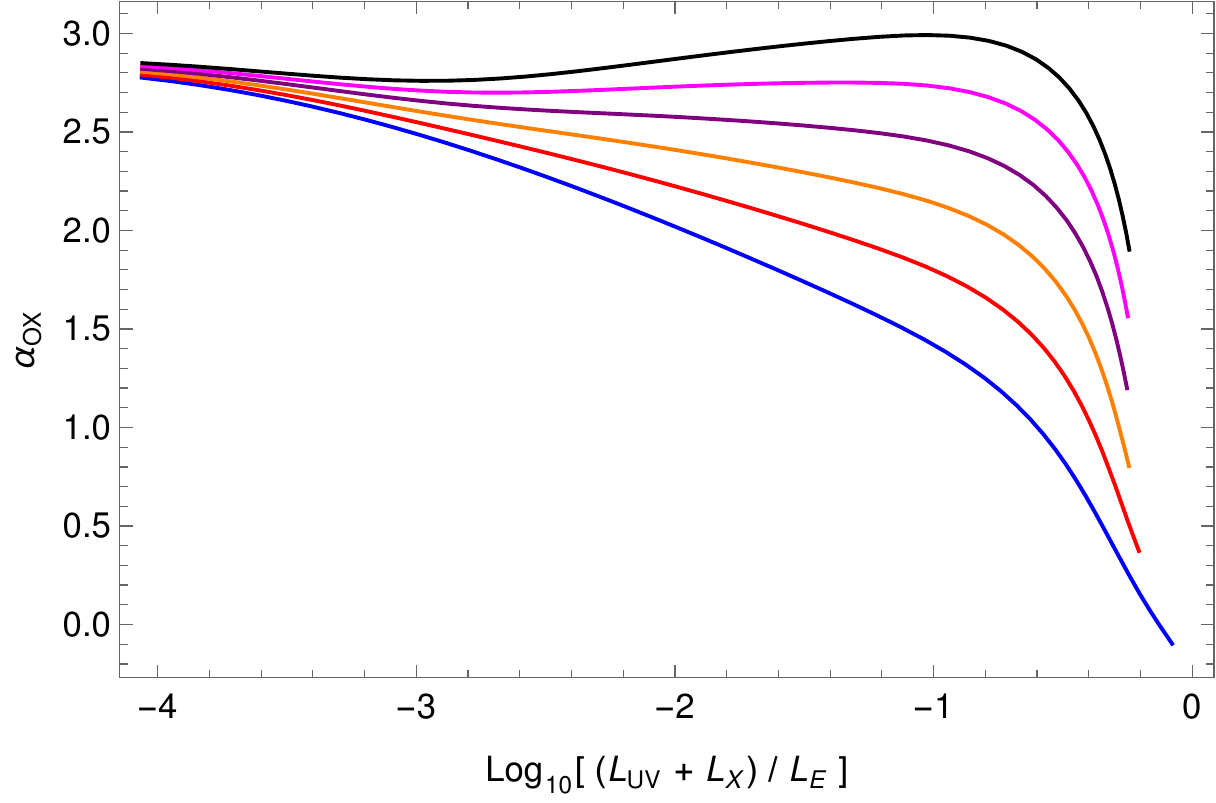}
\end{center}
\caption{\label{mucomp} The evolution of spectral index $\alpha_{\text{OX}}$ with luminosity for relativistic model for the source XMMSL1J0740. The blue, red, orange, purple, magenta and black lines corresponds to $\mu = $ 0, 0.2, 0.4, 0.6, 0.8 and 1 respectively. The other parameters are obtained from Table \ref{rres}. See section \ref{discuss} for details.}
\end{figure}

We have developed a relativistic steady advective accretion model with corona for a viscous stress that is a function of both gas ($P_g$) and total pressure ($P_t$), and given by $\tau_{r\phi} \propto P_g^{\mu} P_t^{1-\mu}$. The small value of obtained $\mu$ implies that the viscous stress is dominated by the total pressure. When the viscous stress is dominated by total pressure (small $\mu$), the spectral index decreases with an increase in luminosity as can be seen from FIG.~\ref{mucomp}. The spectral index $\alpha_{\text{OX}}$ increases with an increase in $\mu$. The increase in $\alpha_{\text{OX}}$ implies that the X-ray luminosity from the corona decreases compared to the UV luminosity from the disk. With a decrease in the mass accretion rate, the disk temperature decreases and thus the disk luminosity. For the lower mass accretion rate, the disk and the total pressure are dominated by the gas pressure, and thus the viscous stress is a function of gas pressure and nearly independent of $\mu$. Thus, the spectral index shows a weak variation with $\mu$ at lower luminosity. With an increase in the radiation pressure, the total pressure deviates from gas pressure, and the effect of $\mu$ emerges.

\begin{figure}[h!]
\begin{center}
\subfigure[XMMSL2J1446]{\includegraphics[scale=0.65]{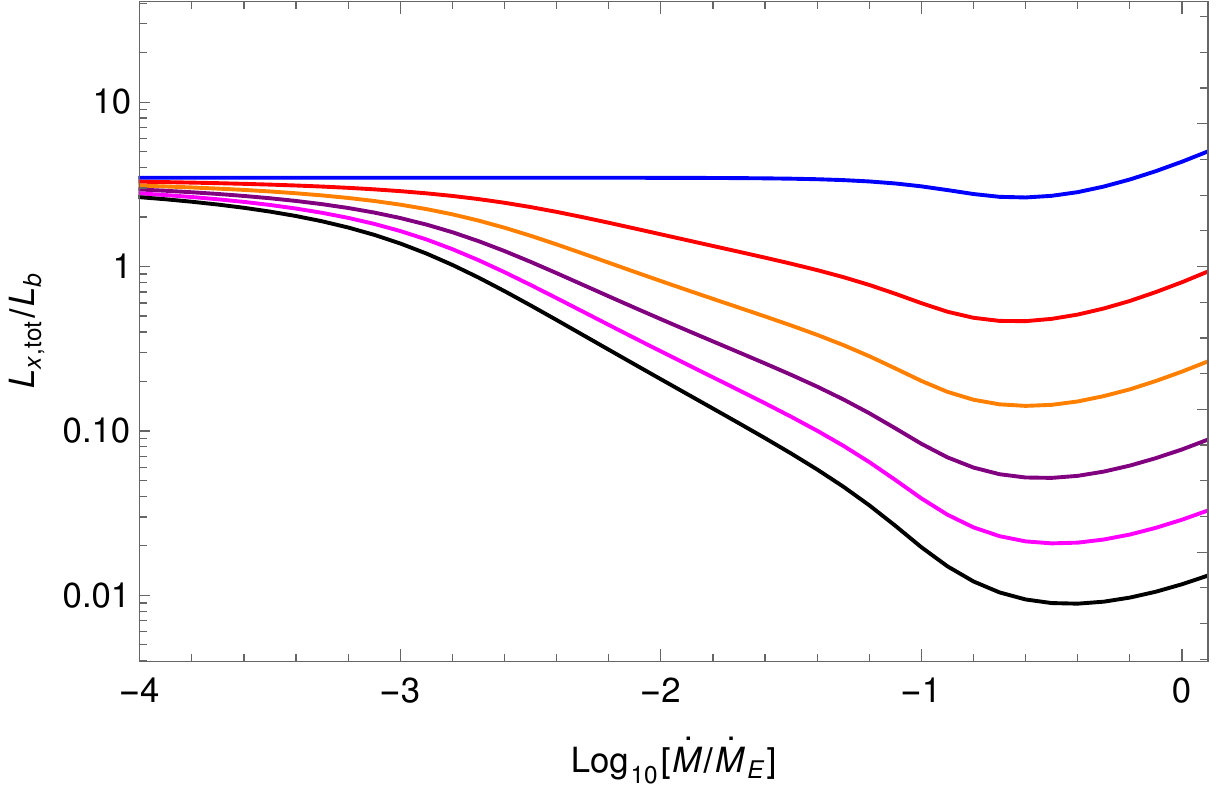}}
\subfigure[XMMSL1J1404]{\includegraphics[scale=0.65]{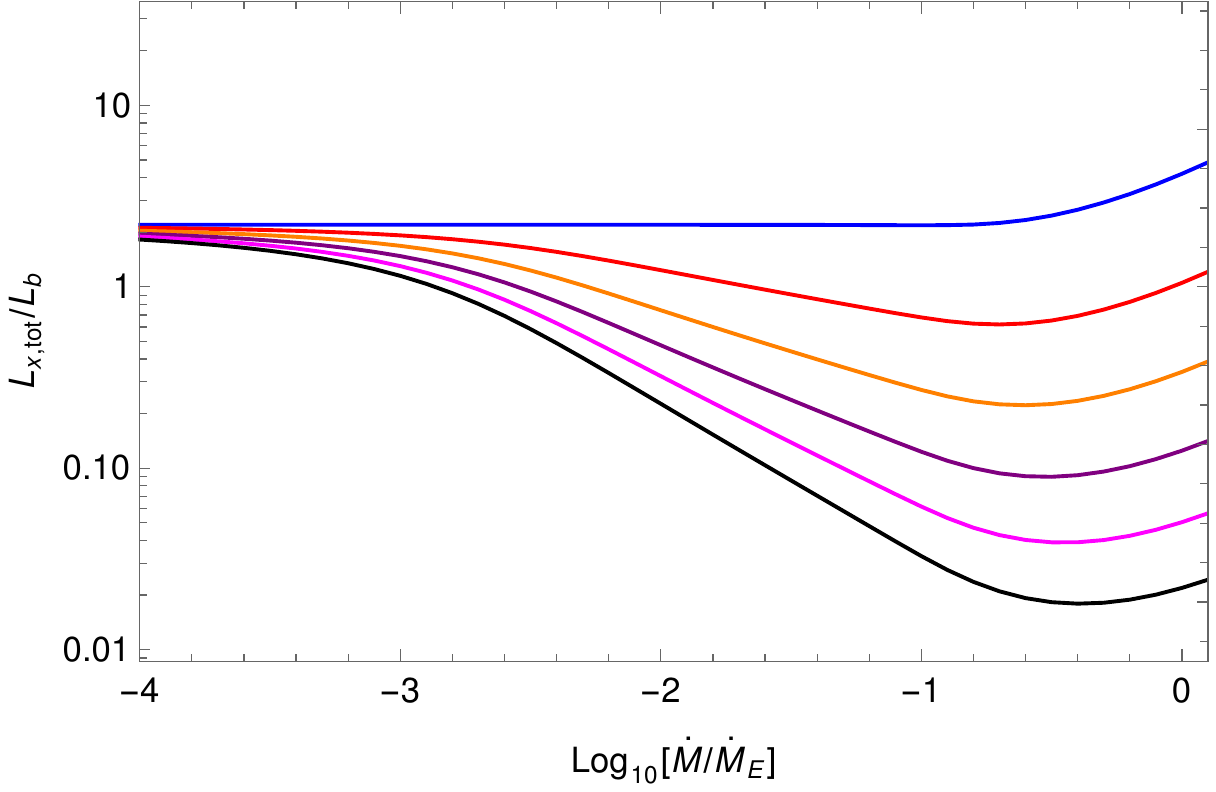}}
\subfigure[XMMSL1J0740]{\includegraphics[scale=0.65]{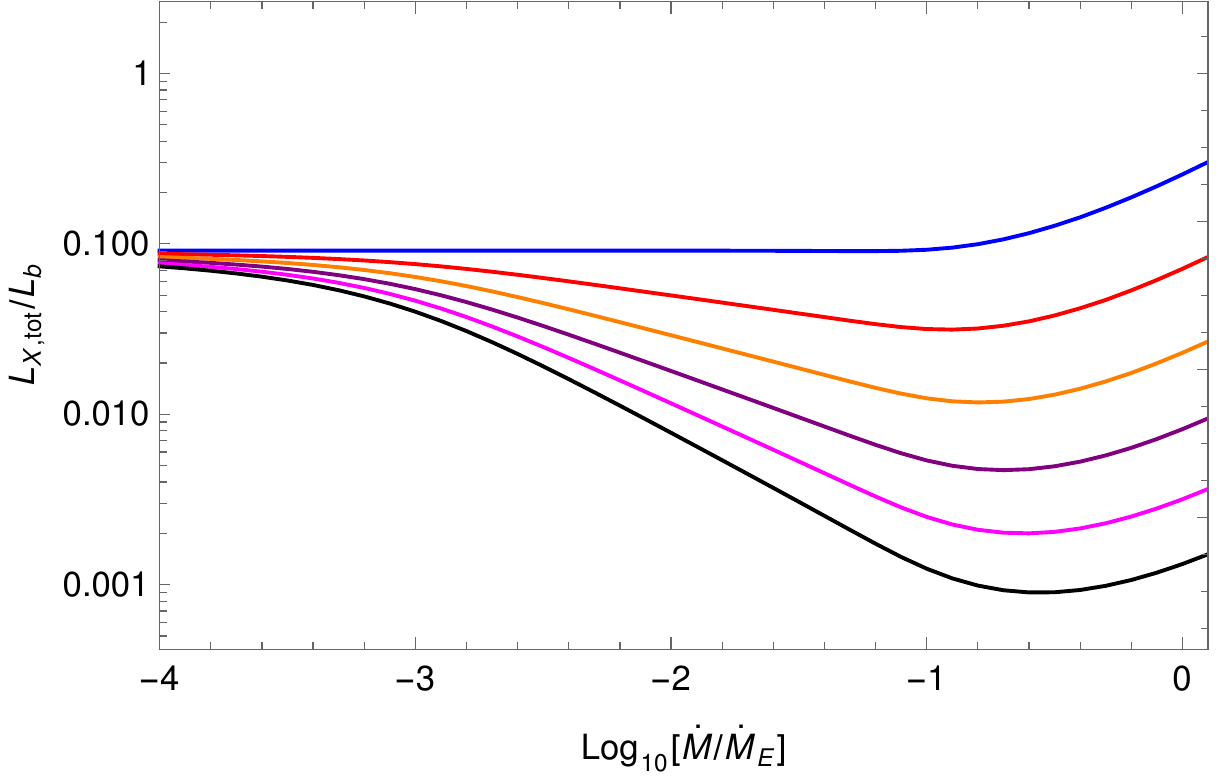}}
\end{center}
\caption{\label{LcLb} The ratio of total X-ray luminosity from the corona to the bolometric disk luminosity for the relativistic model. The blue, red, orange, purple, magenta, and black lines corresponds to $\mu$ = 0, 0.2, 0.4, 0.6, 0.8, and 1. The other parameters in the models have values given in Table \ref{rres}.  See section \ref{discuss} for details.}
\end{figure}

In FIG.~\ref{LcLb}, we can see that for an increase in $\mu$, the total X-ray luminosity from the corona decreases compared to the bolometric luminosity of the disk with mass accretion rate. The MRI growth rate depends on the ratio of radiation to gas pressure and with an increase in $\mu$, the contribution of gas pressure increases compared to the radiation pressure which reduces the MRI and thus the magnetic stress \citep{2002ApJ...566..148T}. This signifies that the gas pressure acts as a counterpoise to the MRI and assists the disk for stability.

\begin{figure}
\begin{center}
\includegraphics[scale=0.7]{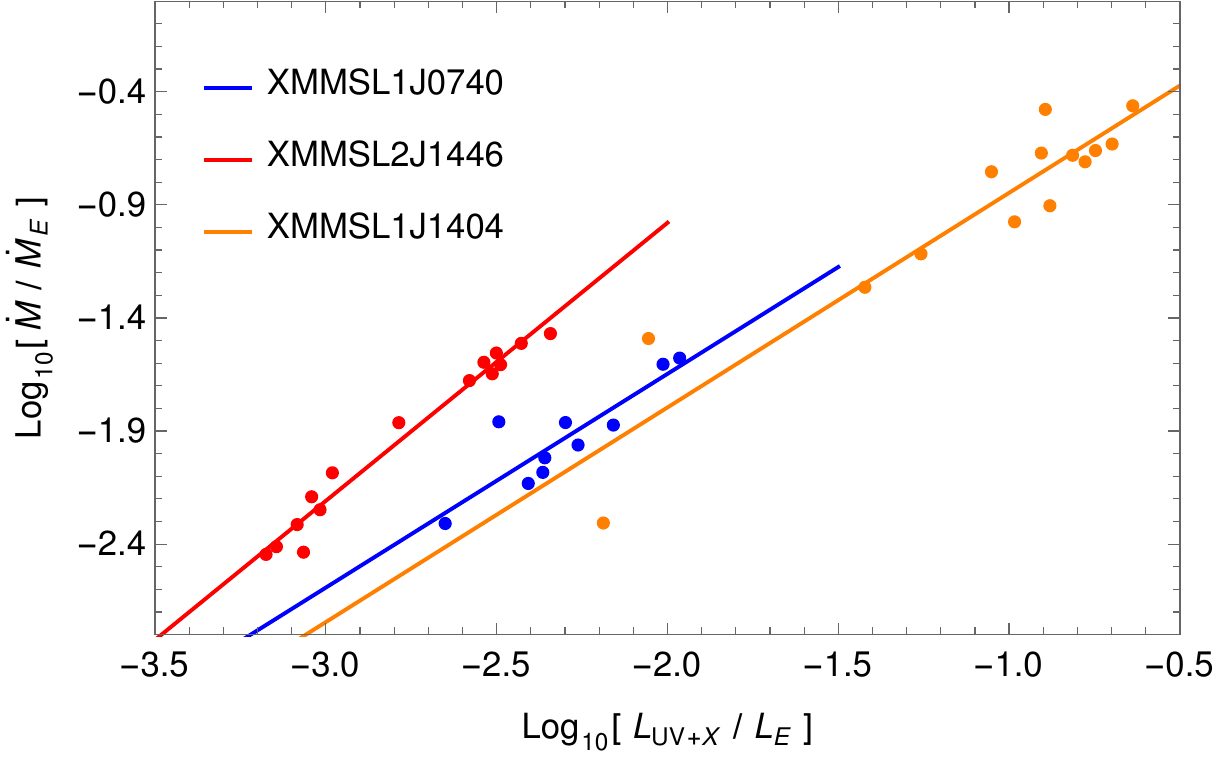}
\end{center}
\caption{\label{mdot} The estimated mass accretion rate corresponding to the individual observed luminosity points for the relativistic models by minimizing the equation (\ref{chi}) at individual points for the parameters given in Table \ref{rres}. The solid lines are linear model fit given by $\log_{10}[\dot{M}/\dot{M}_E] = c_1 + u \log_{10}[(L_X+L_{\text{UV}})/L_E]$. The obtained values are $\{c_1,~u\}$ =  \{0.242, 0.944\} (blue), \{1.478, 1.23\} (red), \{0.102, 0.948\} (orange). See section \ref{discuss} for details. }
\end{figure}

We have constructed the disk-corona plots by considering the mass accretion rate in the range $\dot{M} \in \{10^{-4},~10\}\dot{M}_E$. We estimated a set of parameters for which the models fit the observations. However, the mass accretion rate corresponding to individual observed points is unknown. With the obtained set of parameters, we calculate the mass accretion rate for individual points of the sources. Since the spectral index and luminosity are a function of $\dot{M}$, we perform a minimization of the total variance which is the sum of variance calculated for the spectral index and luminosity for each point. The $\chi^2$ at each point is given by equation (\ref{chi}). We minimize the $\chi^2$ for each points to calculate $\dot{M}$ by taking the other parameter values given in Tables \ref{nrres} and \ref{rres}. The obtained mass accretion rate for the observed luminosity for three sources is shown in FIG.~\ref{mdot}. The mass accretion rate obtained from the relativistic model (FIG.~\ref{mdot}) is higher than the accretion rate obtained from the non-relativistic model (FIG.~\ref{nrmdot}). The mass accretion rate for the source XMMSL1J1404 is close to the Eddington rate. We then fit the obtained mass accretion rate to the observed luminosity by a power-law relation and we found that $(L_X+L_{\text{UV}})/L_E \propto \dot{M}^{u_1}$, where $u_1 = 1/u =$ 0.813, 1.054 and 1.058 for XMMSL2J1446, XMMSL1J1404 and XMMSL1J0740 respectively. Thus, the luminosity $L_X+L_{\text{UV}}$ nearly follows the mass accretion rate similar to the bolometric luminosity from the disk where $L_b \propto \dot{M} c^2$. 

\begin{figure}
\begin{center}
\includegraphics[scale=0.68]{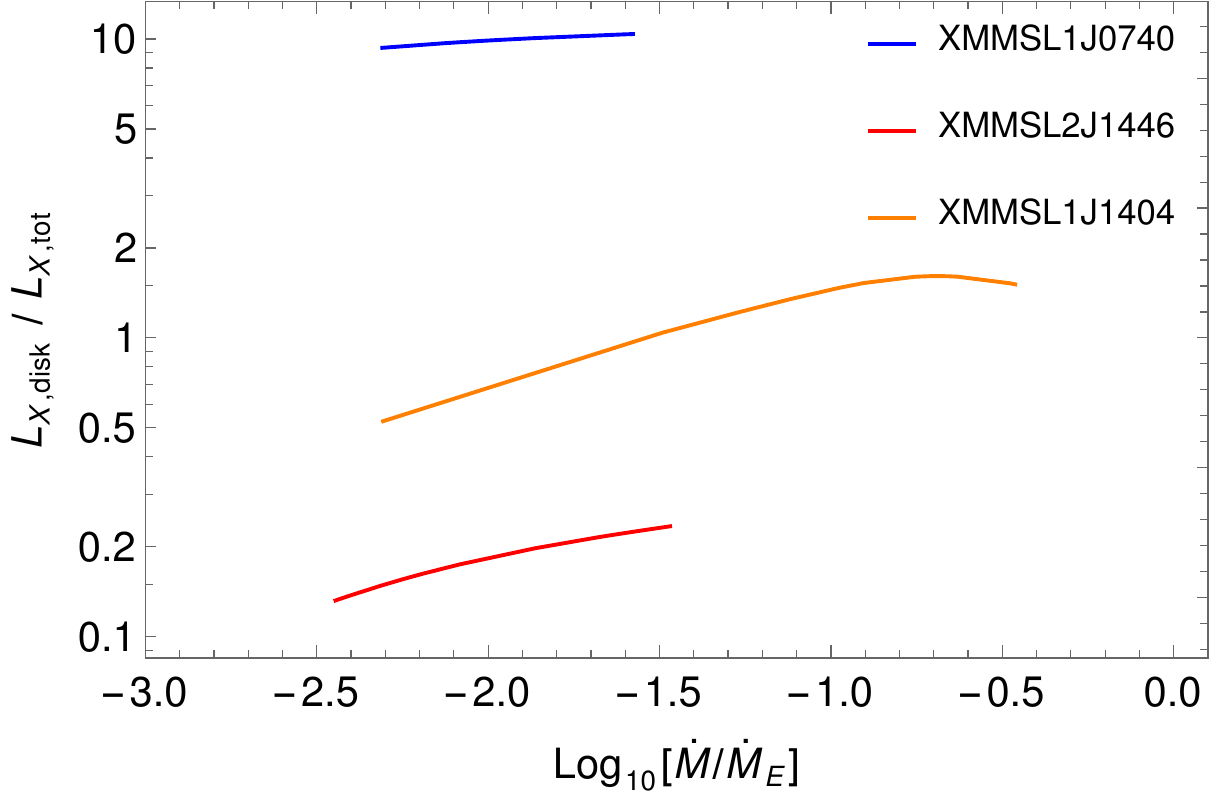}
\end{center}
\caption{\label{ldlcomp} The ratio of X-ray luminosity from the disk ($L_{\text{X,disk}}$) to corona ($L_{\text{X,tot}}$) as a function of obtained mass accretion rate (see FIG.~\ref{mdot}) for the relativistic model. With an increase in $\dot{M}$, the X-ray luminosity from the disk increases and even dominates at a higher accretion rate.  See section \ref{discuss} for details.}
\end{figure}

The ratio of X-ray luminosity from the disk to X-ray luminosity from the corona given by $L_{\text{X,disk}}/L_{\text{X,corona}}$ as a function of obtained $\dot{M}$ is shown in FIG.~\ref{ldlcomp}. The ratio increases with an increase in $\dot{M}$ which implies that the disk contribution increases implying a soft X-ray spectrum for a high mass accretion rate. The ratio exceeds unity for XMMSL1J1404 and XMMSL1J0740 at higher $\dot{M}$. The ratio is higher for the relativistic model (FIG.~\ref{ldlcomp}) compared to the non-relativistic model (FIG.~\ref{nrmdot}) because of the relativistic effects that increase the spectral luminosity as can be seen from FIG.~\ref{J1404comp}. Even though $L_{\text{X,disk}}$ dominates at a higher mass accretion rate, we have shown that the accretion models without corona are unable to explain the observations and the presence of corona is essential. 

The obtained $\mu$ for the sources XMMSL2J1446, XMMSL1J1404 and XMMSL1J0740 implies that the viscous stress is dominated by total pressure. The gas pressure dominates for highly sub-Eddington accretion and the radiation pressure dominates for accretion close to the Eddington rate. For accretion close to Eddington, the total pressure is dominated by radiation pressure which results in the viscous stress that is dominated by radiation pressure. If there is no advection ($Q_{\text{adv}} = 0$) such that viscous heating flux is equal to the radiative flux, the disks with radiation pressure results in a Lightman-Eardley instability \citep{1974ApJ...187L...1L}, where viscous stress $\tau_{r \phi} \propto \Sigma^{-1}$. Thus, the Lightman-Eardley instability is for a disk with radiation pressure and no advection energy loss. However, we have included the advection energy loss in the models to have a consistent evolution through Eddington to sub-Eddington phases. The TDE disk evolves with time and thus, the mass accretion rate changes as can be seen from FIG.~\ref{mdot}. As the mass accretion rate decreases, the total pressure tends to gas pressure and the disk is thermally stable. A detailed analysis of thermal instability in an advective disk will require a time-dependent simulation.

In the relativistic model presented in section \ref{rdcm}, we have approximated the differential element of solid angle $\diff \Theta$  subtended on the observer’s sky by the disk element in the Newtonian limit (see equation \ref{dtheta}) and considered the disk to be face on to the observer such that $\theta_{\text{obs}} = 0^{\circ}$. We have included the relativistic area element of the disk, and the gravitational and Doppler redshift to calculate the emissions. We have relaxed the details of scattering in the corona which includes the synchrotron and bremsstrahlung emissivities that depend on the electron distribution in the corona and approximated the emission using a downward component $\eta$. \citet{2012ApJ...761..109Y} showed using relativistic radiative transfer, a plane parallel corona with non-thermal emissivities and ray-tracing that the spectra of the disk–corona systems vary with inclination angle $\theta_{\text{obs}}$, but the X-ray to bolometric luminosity is insensitive to $\theta_{\text{obs}}$. The ratio of X-ray luminosity to bolometric luminosity declines with mass accretion rate by an order of 10-100 over a mass accretion rate range of $\{10^{-2}-1\} \dot{M}_E$. Our relativistic model applied to TDEs shows a similar result as can be seen from FIG.~\ref{LcLb}. 

The evolution of $\alpha_{\text{OX}}$ and its positive and negative correlations with luminosity is useful to study the accretion state change in the disk-corona system. We have developed steady disk-corona models and applied them to TDEs with an X-ray spectrum dominated by a power law. TDE sources XMMSL2J1446, XMMSL1J1404, and XMMSL1J0740 show a negative correlation between the spectral index and the luminosity as can be seen from FIG.~\ref{fitplt}. Now, we consider a TDE source AT2018fyk which shows a positive correlation between the spectral index and the luminosity. 

\subsection{\label{fyksec} AT2018fyk} 

The transient AT2018fyk was discovered on 2018 September 8 by the All-Sky Automated Survey for Supernovae in the galaxy at a redshift of 0.059. \citet{2021ApJ...912..151W} performed an analysis of both archival and new optical, UV, and X-ray observations of the source taken up to 2 yr after the initial discovery. The spectral index shows a positive correlation with the bolometric luminosity. The black hole mass estimated using the $M_{\bullet}-\sigma$ relation results in $\log_{10}[M_{\bullet}] = 7.7 \pm 0.4 M_{\odot}$, and using the break frequency-black hole mass scaling relation for AGNs \citep{2006Natur.444..730M} results in $\log_{10}[M_{\bullet}] = (6.9–7.2) \pm 0.55 M_{\odot}$. We use the upper limit of mean value given by $\log_{10}[M_{\bullet}] = 7.2 M_{\odot}$ which is close to the lower limit of black hole mass from $M_{\bullet}-\sigma$ relation. The power index at late time obtained by \citet{2021ApJ...912..151W} is be $\Gamma = 2.1 \pm 0.1$.

\begin{figure*}
\begin{center}
\subfigure[]{\includegraphics[scale=0.7]{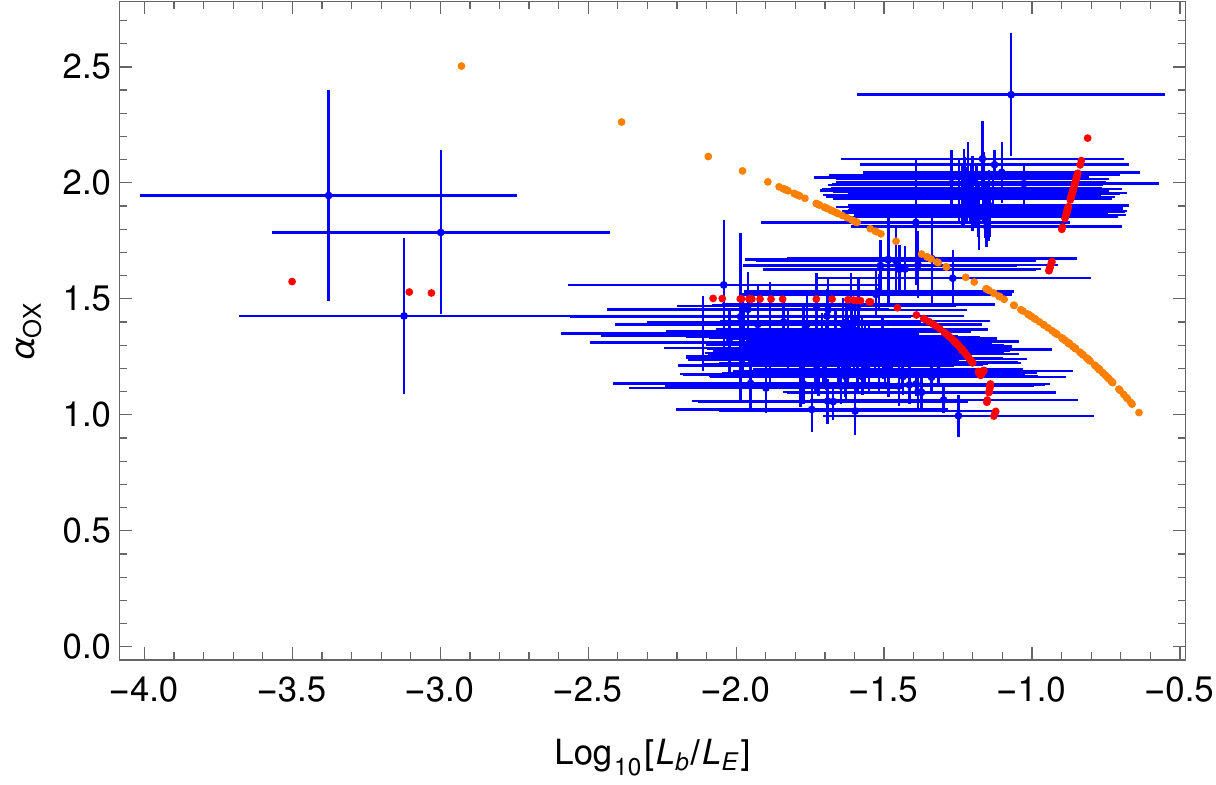}}
\subfigure[]{\includegraphics[scale=0.7]{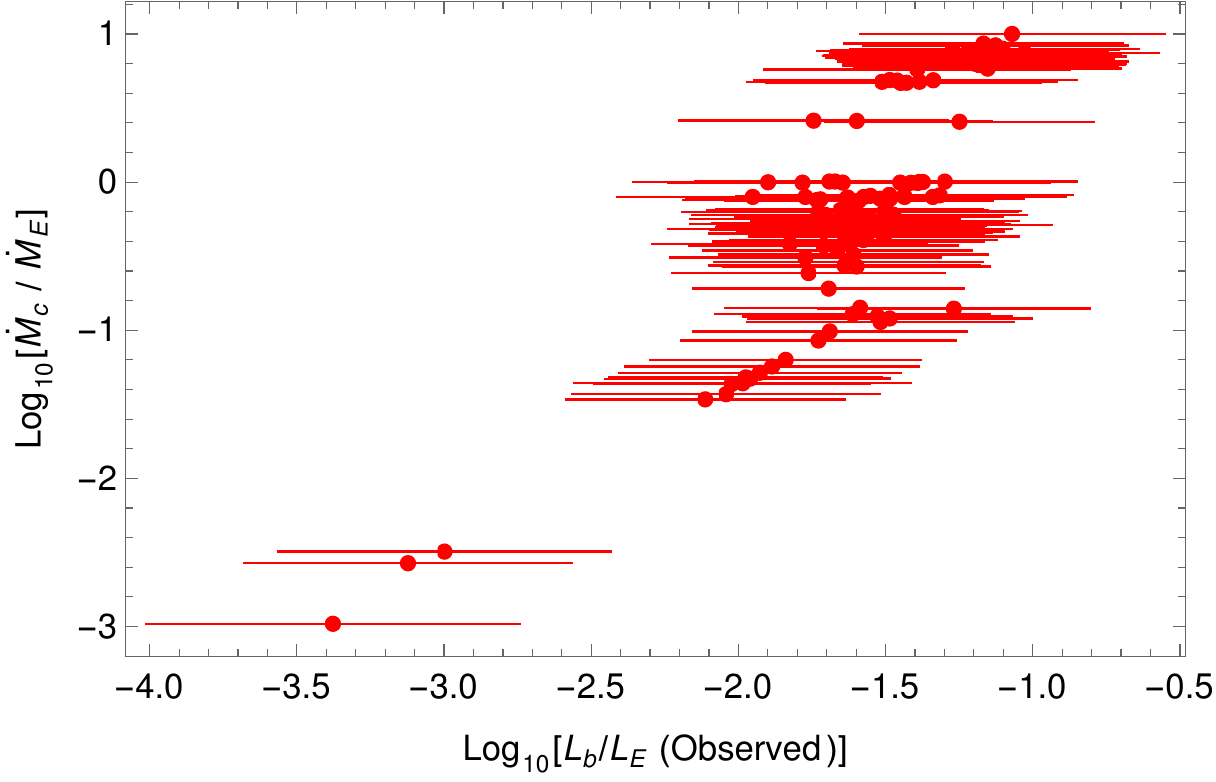}}
\end{center}
\caption{\label{atfyk} (a) The spectral index $\alpha_{\text{OX}}$ obtained using relativistic disk-corona model (section \ref{rdcm}) with spherical outflow wind (red points) and without spherical outflowing wind (orange points). The blue points are the observational data taken from \citet{2021ApJ...912..151W}. We can see that an outflow wind is required for the positive correlation between the spectral index and luminosity. (b) The total mass loss rate was obtained using the equation (\ref{chi}) for the relativistic disk-corona model with a spherical outflow. See section \ref{fyksec} for details. }
\end{figure*}

We apply the relativistic disk-corona model to the observation and it does not provide a good fit as can be seen from the orange points in FIG.~\ref{atfyk}. To explain the observed spectral index, we introduce an outflowing wind to the relativistic disk-corona model presented in section \ref{rdcm}. The relativistic disk-corona model remains true except the mass accretion rate $\dot{M} = (1- f_{\text{out}}) \dot{M}_c$, where $\dot{M}_c$ is total mass loss rate such that the mass outflow rate is $\dot{M}_{\text{out}} = f_{\text{out}} \dot{M}_c$. The disk and corona solution is obtained using the equations given in section \ref{rdcm} with $\dot{M} = (1- f_{\text{out}}) \dot{M}_c$. To estimate the outflow structure, we assume the outflow to be spherical and adiabatic with Thomson opacity. The outflow is launched from the radius $r_L = r_{\text{out}}$, where $r_{\text{out}}$ is disk outer radius. \citet{2009MNRAS.400.2070S} constructed an adiabatic spherical model and obtained the radius and the temperature of the photosphere of the wind given by 

\begin{eqnarray}
r_{\text{ph}} &=& \frac{f_{\text{out}}\dot{M}_a\kappa_{\rm es}}{4\pi v_{w}} \\ 
T_{\text{ph}} &=& (4\pi)^{\frac{5}{12}}\left(\frac{1}{2a}\right)^{\frac{1}{4}}\kappa_{\text{es}}^{-\frac{2}{3}} \frac{f_{\text{ out}}^{-\frac{5}{12}}f_{v}^{\frac{11}{12}}}{\dot{M}_c^{\frac{5}{12}}}\frac{(GM_{\bullet})^{\frac{11}{24}}}{r_{L}^{\frac{7}{24}}}
\end{eqnarray}

where $v_w=f_{v}\sqrt{G M_{\bullet}/r_L}$ is the velocity of the outflowing wind with $f_{v}$ taken to be unity and the fraction of mass outflow $f_{\text{out}}=\dot{M}_{\text{out}}/\dot{M}_c$ is given by \citep{2015ApJ...814..141M}
 
\begin{equation}
f_{\text{out}}=\frac{2}{\pi} \arctan \left[\frac{1}{4.5}\left(\frac{\dot{M}_c}{\dot{M}_E}-1\right)\right].
\end{equation}

\noindent The bolometric luminosity of the outflowing wind is $L_b^w=4\pi r_{ph}^2 \sigma_{SB} T_{ph}^4$. We assume that the blackbody UV emission is from the wind when the accretion rate exceeds the Eddington rate else the UV emission is from the disk, and the X-ray emission is from the disk and the corona. The obtained parameters are shown in Table \ref{fyktab} and the obtained spectral index is shown in red points in FIG.~\ref{atfyk}. By including the outflow, a positive correlation is obtained between the spectral index and the luminosity. The total mass accretion rate $\dot{M}_c$ exceeds the Eddington rate for higher luminosity and thus suggests the presence of an outflowing wind.

The outflowing wind due to a strong radiation pressure in the super-Eddington disk results in an emission that dominates in the optical/UV bands \citep{2011MNRAS.410..359L}. This results in an enhancement in the UV luminosity. The X-ray emission from the wind is weak due to the photosphere temperature that is smaller than the disk single blackbody temperature ($\sim 10^{5}~{\rm K}$) \citep{2020SSRv..216..114R}. The disk-wind model is used to explain the TDE observations where optical/UV emissions dominate over X-ray emissions. The equation (\ref{aox}) results in $\lambda L_{\rm 2500 \AA} / \lambda L_{\rm 2~keV} = (\nu_{\rm 2~keV}/\nu_{\rm 2500 \AA})^{\alpha_{\rm OX} -1} = (401.2)^{\alpha_{\rm OX} -1}$. Thus, an increase in $\alpha_{\rm OX}$ implies an increase in the UV 2500 \AA~  luminosity over the X-ray 2 keV luminosity. The total luminosity is the sum of UV and X-ray luminosity and increases with an increase in the luminosity of X-ray and/or UV. The positive correlation between $\alpha_{\rm OX}$ and bolometric luminosity implies that the UV luminosity increases with an increase in bolometric luminosity.

\begin{figure}
\begin{center}
\subfigure[]{\includegraphics[scale=0.65]{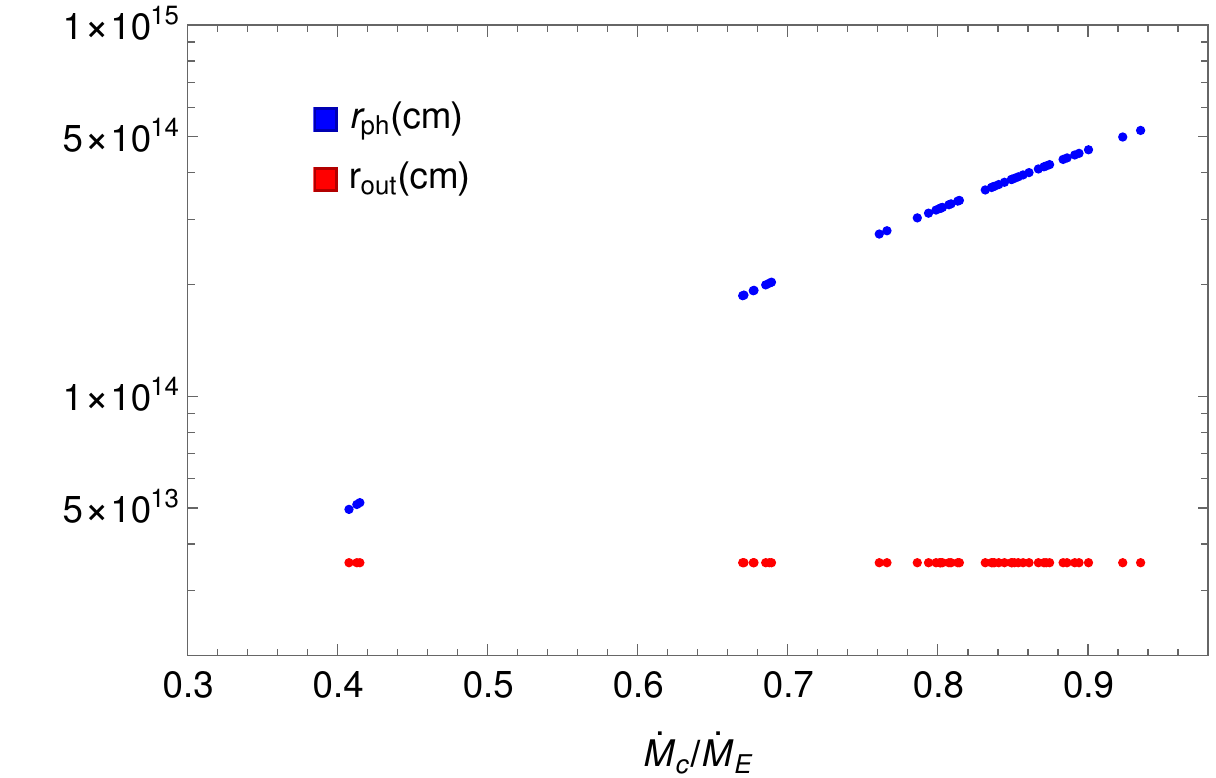}}
\subfigure[]{\includegraphics[scale=0.65]{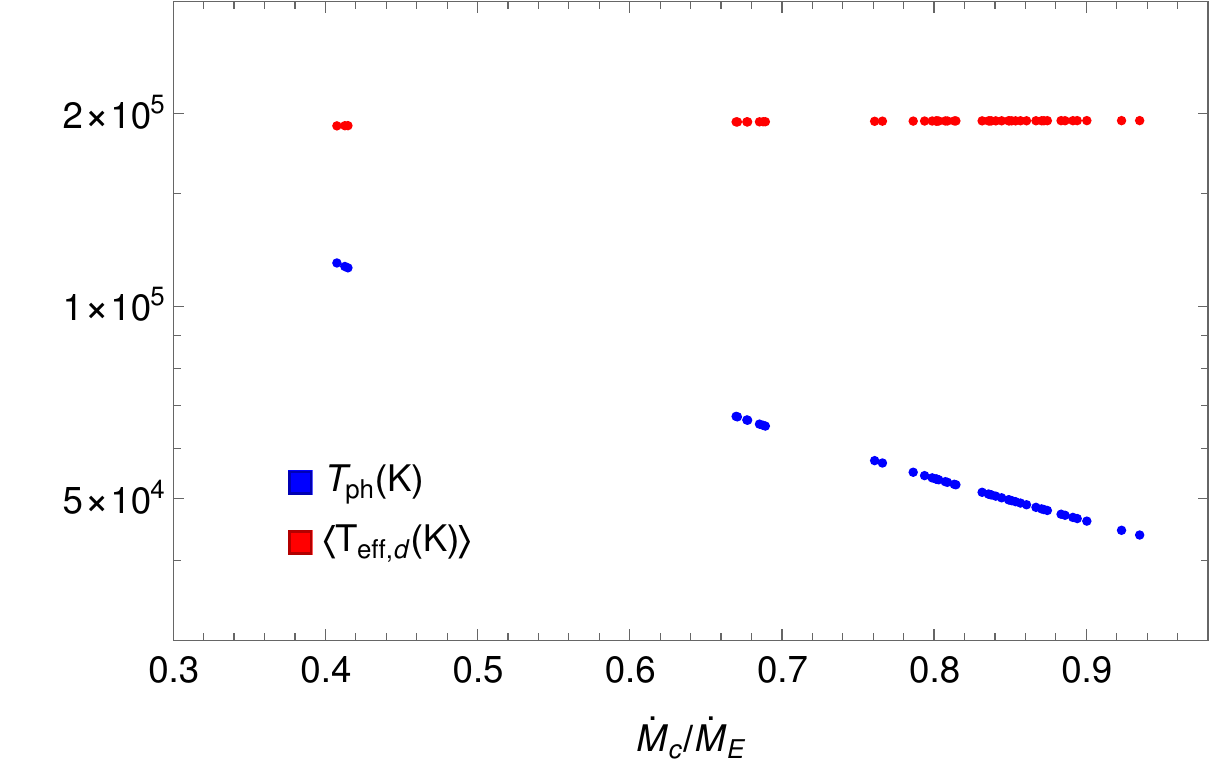}}
\subfigure[]{\includegraphics[scale=0.65]{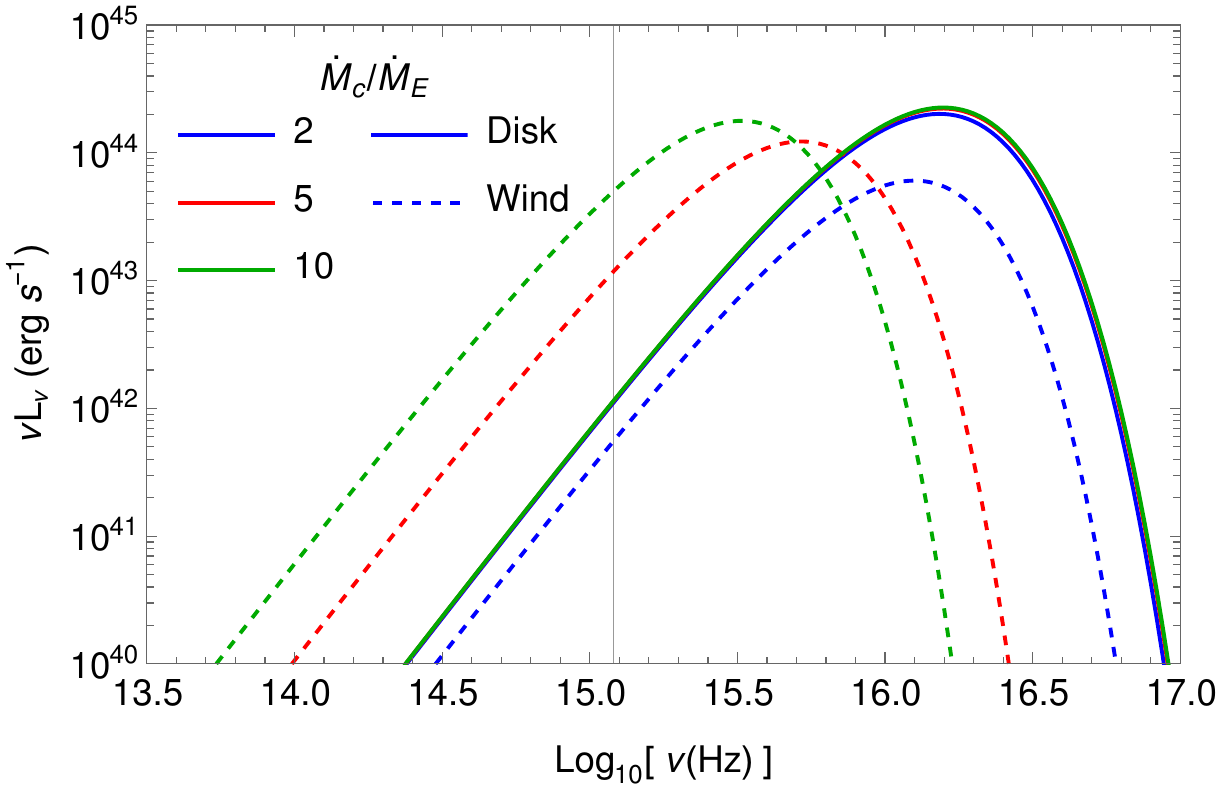}}
\end{center}
\caption{\label{atfyk_p} The evolution of photosphere radius, temperature, and spectrum from the disk-wind model with mass accretion rate for AT2018fyk are shown. The parameters are given in Table \ref{fyktab}. (a) The photosphere radius of wind ($r_{\rm ph}$) emission is higher than the disk outer radius ($r_{\rm out}$) and increases with mass accretion rate. (b) The photosphere temperature ($T_{\rm ph}$) is lower than the radial averaged disk temperature and decreases with mass accretion rate. (c) The spectrum from wind emission is shown by dashed lines whereas the disk spectrum is shown by solid lines. The spectrum from the wind emission dominates in the optical/UV bands and is higher than the disk luminosity. The X-ray luminosity from the wind is negligible for a higher mass accretion rate. The 2500 \AA~ line is shown by the vertical grey line.  See section \ref{fyksec} for details.}  
\end{figure}

FIG.~\ref{atfyk_p} shows the photosphere radius, temperature and spectrum for the wind emission. The photosphere radius increases, whereas the photosphere temperature decreases with the mass accretion rate. The photosphere temperature is lower than the radial averaged disk temperature ($\left<T_{\rm eff,d}\right> = \int T_{\rm eff} \, \diff \mathcal{A}/ \int \, \diff \mathcal{A}$, where $T_{\rm eff}$ and $\diff \mathcal{A}$ are disk effective temperature and disk area element). The spectrum from the wind emission dominates in the optical/UV bands with an insignificant X-ray emission, and the UV luminosity increases with the mass accretion rate. The enhancement in the UV luminosity compared to the X-ray luminosity in the disk-wind model increases the bolometric luminosity and the spectral index leading to a positive correlation.  

The presence of a corona around a disk that has an outflowing wind is uncertain, because the strong wind may ruin the low-density corona. If the wind destroys the low-density corona, then there will be no coronal emission, and the spectrum will be dominated by the thermal blackbody emission. With a decrease in the wind strength, the corona will tend toward a stable structure and its non-thermal emission will increase. In the sub-Eddington phase, the X-ray spectrum will be dominated by a non-thermal emission. A time-dependent accretion model with an outflow and the dynamics of the corona around such a disk will require detailed numerical calculations.

\begin{table}
\caption{\label{fyktab} The Parameter values are obtained by minimizing the equation (\ref{chi}) using the relativistic disk-corona model along with the reduced chi-square ($\chi_r^2$) for TDE source AT2018fyk. The model fit to the observation is shown in FIG.~\ref{atfyk}.}
\begin{ruledtabular}
\begin{tabular}{ccc}
&&\\
Parameter & Relativistic disk-corona & Relativistic disk-corona  \\
& Without outflow & With outflow \\
&&\\
\hline
&&\\
$\chi_r^2$ & 2.4 & 0.87 \\
&&\\
$\alpha_0$ & 0.179 & 0.437 \\
&&\\
$\mu$ & 0.12 & 0.595 \\ 
&&\\
$\eta$ & 0.829 & 0.0032 \\
&& \\
$a_d$ & 0.262 & 0.902 \\
&&\\
$M_{\star}(M_{\odot})$ & 1.094 & 1.0087 \\
&&\\
$j$ & 0.672 & 0.141 \\
&&\\
$q$ &  1.00 & 1.02 \\
&&\\
\end{tabular}
\end{ruledtabular}
\end{table}

We have considered a highly simplified spherical outflow model which is a non-relativistic model. The outflow from an accretion disk and the emission from the wind is complex in dynamics, and require a detailed numerical analysis. We used this simple model to show that an outflow is required to explain the positive correlation, else the disk-corona can explain the negative correlation only. 
The caveat of steady accretion models to the TDEs is that the TDE disk evolves with time and thus, requires a time-dependent disk-corona evolution. Such a model will be able to explain the long-term evolution of the TDE disk and the accretion state change inferred from the spectral index evolution with luminosity.   

\section{\label{summary} Summary}

We have developed a steady relativistic disk-corona model with a stress tensor that is assumed to be dominated by Maxwell stresses ($\tau_{r\phi} \propto P_{\text{mag}}$) with magnetic pressure given by $P_{\text{mag}} \propto P_g^{\mu}P_t^{1-\mu}$. In the flux calculation, we have included the redshift factor ($g$) that considers the gravitational and Doppler redshifts. 

The accretion models without corona is unable to explain the observed spectral index for TDE sources (see Figs. \ref{ssd}, \ref{sq}, \ref{can} and \ref{MM}). However, the spectral index calculated from the disk-corona models matches with the observations and thus exhibits the importance of corona. We present the relativistic disk-corona model in section \ref{rdcm}. We have also shown the non-relativistic disk-corona model in appendix \ref{ndcm}. The presence of corona reduces the disk spectrum as can be seen from the non-relativistic model spectrum in FIG.~\ref{J1404comp}. The relativistic dynamics increase the disk spectrum and the luminosity.

We have included the advection energy loss in the energy conservation equation in both the models which is crucial when the accretion is near Eddington where the pressure is dominated by radiation pressure. 

We fit the relativistic model to the three TDE sources and the parameters value are given in Table \ref{rres}. The parameter $\mu$ plays a significant role in the luminosity ratio $L_{\text{ X, tot}}/L_b$ with an increase in the mass accretion rate as can be seen from FIG.~ \ref{LcLb}. The estimated value of $\mu$ for the sources XMMSL2J1446, XMMSL1J1404 and XMMSL1J0740 are $\mu =$ $\sim$0, 0.203 and $\sim$0 (relativistic). Thus, the stress is dominated by total pressure. With an increase in $\mu$, the spectral index increases, and the luminosity ratio decreases with the mass accretion rate implying the dominance of the disk over the corona emission. 

The X-ray luminosity of corona to disk bolometric luminosity $L_{\text{ X, tot}}/L_b$ is higher for the relativistic model suggesting that the relativistic effects increase the energy transport to the corona. The ratio of X-ray luminosity from the disk to corona increases with an increase in the mass accretion rate (see FIG.~\ref{ldlcomp}). 

We have estimated the mass accretion rate for the individual data points for all the sources by minimizing the equation (\ref{chi}) and the obtained values are shown in Fig. \ref{mdot}, which follows a nearly linear relation with the observed luminosity for both the models. Thus, the luminosity $L_X+L_{\text{UV}}$ follows the mass accretion rate similar to the disk bolometric luminosity where $L_b \propto \dot{M} c^2$.

The observed spectral index with luminosity for TDE AT2018fyk shows a better fit with the relativistic disk-corona model when a simple spherical outflow emission model is included. This is due to the enhanced UV emission from the outflowing wind due to a large photosphere (FIG.~\ref{atfyk_p}). The disk-corona model with an outflow shows a positive correlation when the total mass accretion rate exceeds the Eddington rate.  

%A detailed time-dependent disk inflow-outflow with the corona model will provide a detailed understanding of the spectral index evolution.  
%A non-relativistic model presented in appendix \ref{ndcm} is for comparison and study of the impact of relativistic dynamics on the disk-corona solution and the estimation of the parameters. The obtained parameters show significant variations. 

The steady disk-corona models are capable to explain the spectral index-luminosity observations for TDEs and this motivates to develop a time-dependent disk-corona emission. The time-dependent model will be useful to explain the long-term evolution of spectral index, luminosity, and the accretion state transition.

\begin{acknowledgments}
We thank the referee for the constructive suggestions that have improved the paper. MT has been supported by the Basic Science Research Program through the National Research Foundation of Korea (NRF) funded by the Ministry of Education (2016R1A5A1013277 and 2020R1A2C1007219). 
\end{acknowledgments}

\bibliography{ref}

\appendix

\section{\label{armp} Relativistic disk model equations}

Here, we present the relativistic disk equations formulated by \citet{2020MNRAS.496.1784M} in the cylindrical coordinates. The space-time metric in the geometrical units ($c=G =1$) with the signature ($- + + +$), is given by 

\begin{widetext}
\begin{multline}
\diff S^2 = -\left[1-\frac{2 M (r^2+z^2)^{3/2}}{(r^2+z^2)^2+a^2 z^2}\right] \diff t^2 - \frac{4 M a r^2 \sqrt{r^2+z^2}}{(r^2+z^2)^2+a^2 z^2} \diff t \diff \phi + \frac{(r^2+z^2)^2+a^2 z^2}{(r^2+z^2)^2}\left[\frac{r^2}{r^2+z^2- 2 M \sqrt{r^2+z^2} +a^2}+ \right. \\ \left.\frac{z^2}{r^2+z^2}\right] \diff r^2 + \frac{(r^2+z^2)^2+a^2 z^2}{(r^2+z^2)^2} \left[\frac{z^2}{r^2+z^2- 2 M \sqrt{r^2+z^2} +a^2}+ \frac{r^2}{r^2+z^2}\right] \diff z^2 +  \frac{2 r z [(r^2+z^2)^2+a^2 z^2]}{(r^2+z^2)^2} \\  \left[ \frac{1}{r^2+z^2- 2 M \sqrt{r^2+z^2} +a^2}- \frac{1}{r^2+z^2}\right]  \diff r \diff z +  \frac{r^2}{r^2+z^2} \left[r^2+z^2+a^2 + \frac{2 M a r^2 \sqrt{r^2+z^2}}{(r^2+z^2)^2+a^2 z^2}\right] \diff \phi^2.
\end{multline}
\end{widetext}

In the limit of thin disk $z \ll r$, the metric tensors are given by 

\begingroup
\allowdisplaybreaks
\begin{eqnarray}
g_{\text{t t}} &=& -1+ \frac{2 M }{r}- \frac{M (2 a^2+ r^2) z^2}{r^5}, \\ 
g_{\text{t r}} &=& g_{\text{ r t}} = 0, \\
g_{t \phi} &=& g_{\phi t} =  -\frac{2 M a}{r} + \frac{M (2 a^3 + 3 a r^2)z^2}{r^5}, \\
g_{\text{t z}} &=& g_{\text{z t}} = 0, \\
g_{\text{r r}} &=& \frac{r^2}{r^2 -2 M r + a^2} + \nonumber \\
 &&  \frac{[2 a^4 + 3 a^2 (r^2-2M r)+ M (4 M r^2-3 r^3) ] z^2}{r^2(r^2- 2 M r +a^2)^2}, \label{grr} \\
g_{r \phi} &=& g_{\phi r} = 0, \\
g_{\text{r z}} &=& g_{\text{z r}} = 2 \left(-1+ \frac{r^2}{r^2 - 2 M r + a^2}\right) \frac{z}{r}, \label{gpp}\\
g_{\phi \phi} &=& \frac{r^4+ a^2 r^2 + 2 M a^2 r}{r^2} - \nonumber \\ 
&&  \frac{a^2(r^4 + M r (2 a^2 + 5 r^2)) z^2}{r^6}, \\
g_{\phi z} &=& g_{z \phi} = 0, \\
g_{\text{z z}} &=& 1 + \left[\frac{a^2- r^2}{r^2} + \frac{r^2}{r^2- 2 M r + a^2} \right] \frac{z^2}{r^2},
\end{eqnarray}%
\endgroup

\noindent which is the same as the metric tensor given in \citet{2015PhyU...58..527Z}. At the equatorial plane ($z=0$), the space-time metric reduces to 

\begin{equation}
\diff S^2 = - \left(\frac{r- 2 M}{r}\right) \diff t^2 - \frac{4 M a }{r} \diff t \diff \phi + \frac{r^2}{\Delta} \diff r^2 + \frac{A}{r^2} \diff \phi^2 + \diff z^2,
\end{equation}

\noindent where $\Delta = r^2- 2 M r +a^2$ and $A= r^4+ a^2 r^2 + 2 M a^2 r$.

The mass conservation equation is given by

\begin{equation}
\frac{\partial}{ \partial t}(\Sigma u^t) + \frac{1}{r} \frac{\partial}{\partial r} (r \Sigma u^r) = 0,
\label{massr}
\end{equation} 

\noindent where $\Sigma$ is the surface density, $r$ is the radial coordinate and the covariant four velocities near the equatorial plane are then given by

\begin{eqnarray}
u_t &=& - \frac{\gamma_L r \Delta^{1/2}}{A^{1/2}} -\omega \mathcal{L}, \label{utl}\\
u_r &=&  \frac{r}{\Delta^{1/2}} \frac{V}{\sqrt{1-V^2}}, \label{url}\\
u_{\phi} &=& \mathcal{L}, \label{upl}\\
u_z &=& 0. \label{uzl}
\end{eqnarray}

\noindent The $V$ is the radial velocity in the co-rotating frame, and $\mathcal{L}$ is the angular momentum per unit mass. The Lorentz factor $\gamma_L$ near the equatorial plane is given by 

\begin{equation}
\gamma_L^2 = \frac{1}{1-V^2} + \frac{r^2 \mathcal{L}^2}{A}.
\end{equation}

The angular momentum conservation is given by 

\begin{equation}
\Sigma \left[ \frac{\gamma_L A^{1/2}}{r \Delta^{1/2}} \frac{\partial \mathcal{L}}{\partial t} + \frac{V}{\sqrt{1-V^2}} \frac{\Delta^{1/2}}{r} \frac{\partial \mathcal{L}}{\partial r} \right] + \frac{1}{r} \frac{\partial}{\partial r} (r \bar{S}_{\phi}^r) = 0,
\label{azir}
\end{equation}

\noindent where $\bar{S}_{\phi}^r$ is the vertically integrated viscous stress given by 

\begin{equation}
\bar{S}_{\phi}^r = - \nu \Sigma \frac{\Delta^{1/2} A^{3/2} \gamma_L^3}{r^5 } \frac{\partial \Omega}{\partial r}.
\label{sphir}
\end{equation}

Assuming the vertical fluid velocity to be zero, the conservation equation in the vertical $z$ direction results in 

\begin{equation}
\frac{\gamma_L^2}{\Delta A} \frac{z}{r^7} \Lambda_1 + \frac{2 \gamma_L \mathcal{L}}{\Delta^{1/2} A^{3/2}} \frac{z}{r^4} \Lambda_2 + \frac{\mathcal{L}^2}{A^2} \frac{z}{r} \Lambda_3 + \frac{1}{\rho} \frac{\partial P}{\partial z} =0,
\label{zeq}
\end{equation}

\noindent where 

\begin{eqnarray}
\Lambda_1 &=& M A^2 [4 a^2 + r (r- 4 M)] - 4M^2 a^2 r A [4 a^2 + \nonumber \\ 
&& r (3 r - 4M)] + 4 M^2 a^2 r^2 [4 a^4 M + 4 M r^4 - \nonumber \\ 
&& a^2 r (4 M^2 -5 M r + r^2)], \label{Lambda1} \\
\Lambda_2 &=& - a M A [4 a^2 + r (3 r - 4M)]+ 2 M a r [4 a^4 M + \nonumber \\
&& 4 M r^4 - a^2 r (4 M^2-5 M r + r^2)], \label{Lambda2} \\
\Lambda_3 &=& 4 a^4 M +4 M r^4 -a^2 r (4 M^2-5 M r + r^2). \label{Lambda3}
\end{eqnarray}

Following the vertical integration of equation (\ref{zeq}), the height of the disk is given by

\begin{equation}
\left(\frac{H}{r}\right)^2 = \frac{P}{\rho} \frac{1}{r^2} \left[\frac{\gamma_L^2}{\Delta A} \frac{\Lambda_1}{r^7} + \frac{2 \gamma_L \mathcal{L}}{\Delta^{1/2} A^{3/2}} \frac{\Lambda_2}{r^4} + \frac{\mathcal{L}^2}{A^2} \frac{\Lambda_3}{r}\right]^{-1}. 
\label{height}
\end{equation}

The inner radius is taken to be the innermost stable circular orbit given by \citep{1972ApJ...178..347B}

\begin{equation}
r_{in} = r_{ISCO} = r_g x_{in},
\label{rint}
\end{equation}

\noindent where $x_{in} = Z(j)$ given by 

\begin{equation}
Z(j) =3+Z_2(j)-\sqrt{(3-Z_1(j)) (3+Z_1(j)+2 Z_2(j))}, \label{zjb}
\end{equation}

\noindent and 

\begin{subequations}
\begin{align}
Z_1(j) &=1+(1-j^2)^{\frac{1}{3}} \left[(1+j)^{\frac{1}{3}}+(1-j)^{\frac{1}{3}}\right]\\
Z_2(j) &=\sqrt{3 j^2+Z_1(j)^{2}}.
\end{align}
\end{subequations} 
  
\section{\label{ndcm} Non-relativistic disk-corona model}

We construct a steady advection disk model with an energy loss to the corona. The fraction of energy dissipated from the disk to the corona at any radius $r$ is given by $f = Q_{\text{cor}}/Q^{+}_{\text{v}}$, where the viscous heating flux generated at any $r$ is given by $Q^{+}_{\text{v}} = (3/2) c_s \tau_{r\phi}$ with sound speed $c_s$ and viscous stress $\tau_{r\phi}$ \citep{2019A&A...628A.135A}. The fraction $f$ using equation (\ref{ptot}) and $Q_{\text{cor}} = v_D P_{\text{mag}}$ is given by 

\begin{equation}
f = \sqrt{\frac{2 \alpha_0}{k_1^2}}\left[1 + \frac{P_r}{P_g}\right]^{-\frac{\mu}{2}},
\end{equation}

\noindent where $k_1 = 3 k_0 / (2 b)$. For simplicity, we take $k_0$ and $b$ to be unity in our calculations. In a steady accretion disk, the mass accretion rate, $\dot{M}$, is a constant and we take the azimuthal velocity to be Keplerian such that the angular momentum conservation results in $\nu \Sigma = \dot{M} s(r)/ (3 \pi)$, where $s(r) = 1 - \sqrt{r_{\text{in}}/r}$, $\nu$ is the viscosity and $\Sigma$ is the surface density. The surface density is given by $\Sigma = 2 H \rho $, where the disk height $H = c_s / \Omega_K$ with sound speed $c_s$ and Keplerian angular velocity $\Omega_K = \sqrt{G M_{\bullet}/r^3}$ \citep{2009MNRAS.400.2070S}. The viscous heating is given by $Q^{+}_{\text{v}} = (9/8) \nu \Sigma \Omega_K^2$ and by comparing it with $Q^{+}_{\text{v}} = (3/2) c_s \tau_{r\phi}$, we obtain

\begin{equation}
\frac{P_g^{\mu}P_t^{3/2-\mu}}{\sqrt{\rho}} = \frac{1}{4 \pi k_0 \alpha_0} \Omega_K^2 \dot{M} s(r).
\label{nss1}
\end{equation}

\noindent The advection accretion loss, $Q_{\text{adv}}$, in a steady state is given by \citep{2002apa..book.....F}

\begin{equation}
Q_{\text{adv}} = \frac{\dot{M} c_s^2}{2 \pi r^2} \frac{4 - 3\beta_g}{\Gamma_3 - 1}\left[ -\frac{r}{T} \frac{\partial T}{\partial r} + \left(\Gamma_3 - 1\right) \frac{r}{\rho} \frac{\partial \rho}{\partial r} \right],
\end{equation} 

\noindent where $\beta_g = P_g/P_t$ is the ratio of gas to total pressure, and   

\begin{equation}
\Gamma_3-1 = \frac{(4 - 3 \beta_g)(\gamma-1)}{\beta_g + 12 (1-\beta_g) (\gamma-1)}
\end{equation}

\noindent and $\gamma$ is the ratio of specific heats for constant pressure to constant volume \citep{1939isss.book.....C}. The radiative flux is given by $Q_{\text{rad}} = 4 \sigma T^4 / (3 \tau)$, where $T$ is the disk mid plane temperature and the opacity $\tau = \tau_{\text{es}} + \tau_{a} = (\kappa_{\text{es}} + \kappa_a)\Sigma$, is the sum of Thomson opacity due to electron scattering and Kramers' opacity due to absorption given above equation (\ref{releq2}). Thus, the energy conservation equation given by equation (\ref{eqnss}) is 

\begin{multline}
\frac{\dot{M} c_s^2}{2 \pi r^2} \frac{4 - 3\beta_g}{\Gamma_3 - 1}\left[ -\frac{r}{T} \frac{\partial T}{\partial r} + \left(\Gamma_3 - 1\right) \frac{r}{\rho} \frac{\partial \rho}{\partial r} \right] = \\ (1 - f [1-\eta(1-a_d)]) \frac{3}{8\pi}\dot{M}s(r) \Omega_K^2 - \frac{4 \sigma T^4}{3 \tau}.
\label{nss2}
\end{multline}

By solving equations (\ref{nss1}) and (\ref{nss2}) with density and temperature zero at the inner radius taken to be the innermost stable circular orbit (ISCO), we obtain the density $\rho$ and the temperature $T$, which is then used to calculate the effective temperature of the disk $T_{\text{eff}} = (Q_{\text{rad}}/\sigma)^{1/4}$ and thus, the luminosity in various spectral bands following a blackbody emission.

We assume that the radiation of the corona is isotropic and that the emissivity is homogeneous in the vertical direction of the corona. The corona is assumed to be plane-parallel in geometry \citep{2009MNRAS.394..207C}. The energy dissipated to the corona at each radius is given by $Q^{\text{cor}} = f  Q^{+}_{\text{v}}$ and a small fraction of this will be seen as the luminosity given by 

\begin{equation}
L_{\text{X,tot}} = \int_{r_{\text{in}}}^{r_{\text{out}}} 2 \pi r (1-\eta) Q^{\text{cor}} \, \diff r.
\end{equation} 

\begin{figure}[h!]
\begin{center}
\includegraphics[scale=0.55]{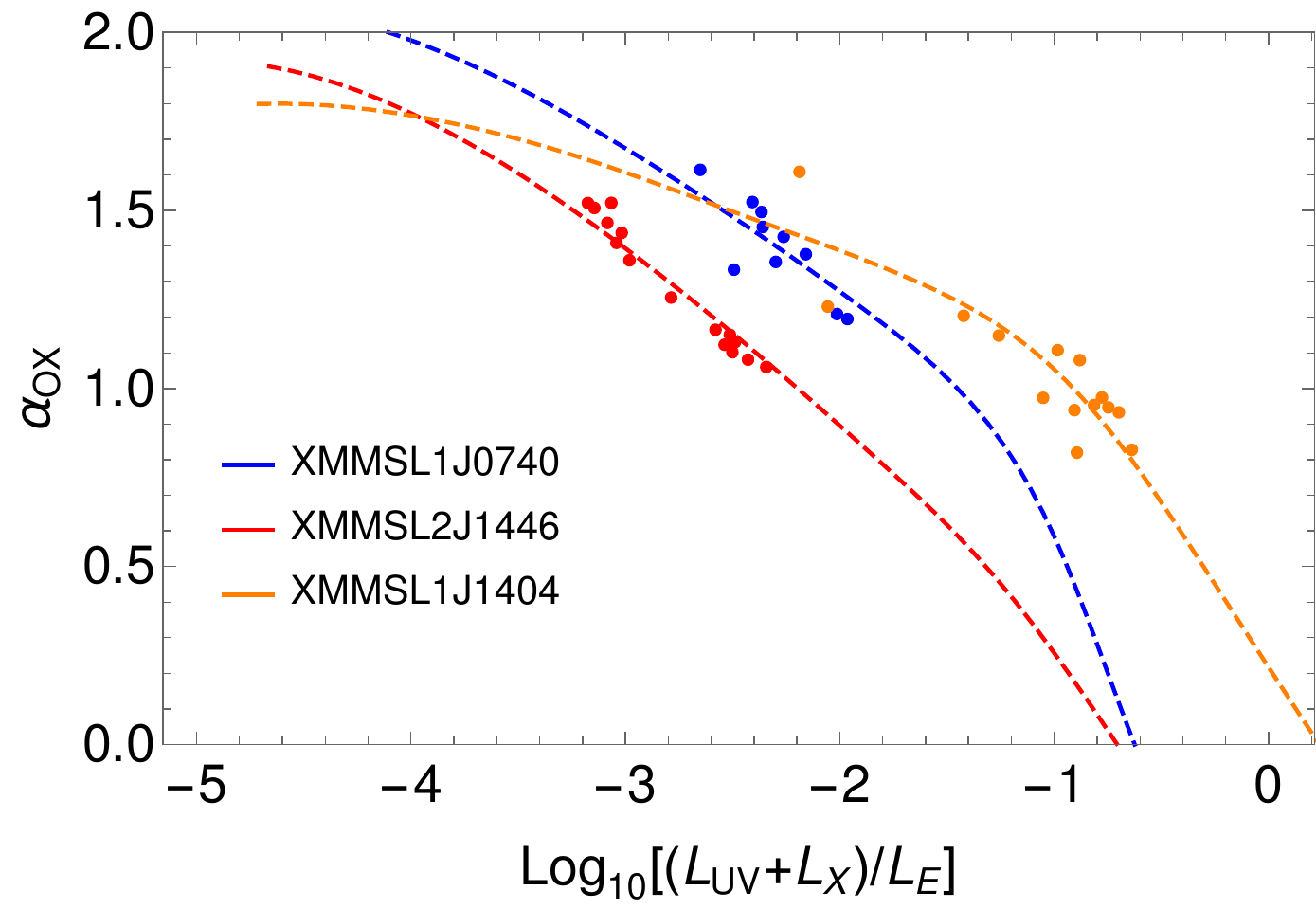}
\end{center}
\caption{\label{fitplt1} The figure shows the non-relativistic model fit to the observations of the considered TDE sources. The obtained values are shown in Tables \ref{nrres}.}
\end{figure}

\begin{figure*}
\begin{center}
\subfigure[]{\includegraphics[scale=0.7]{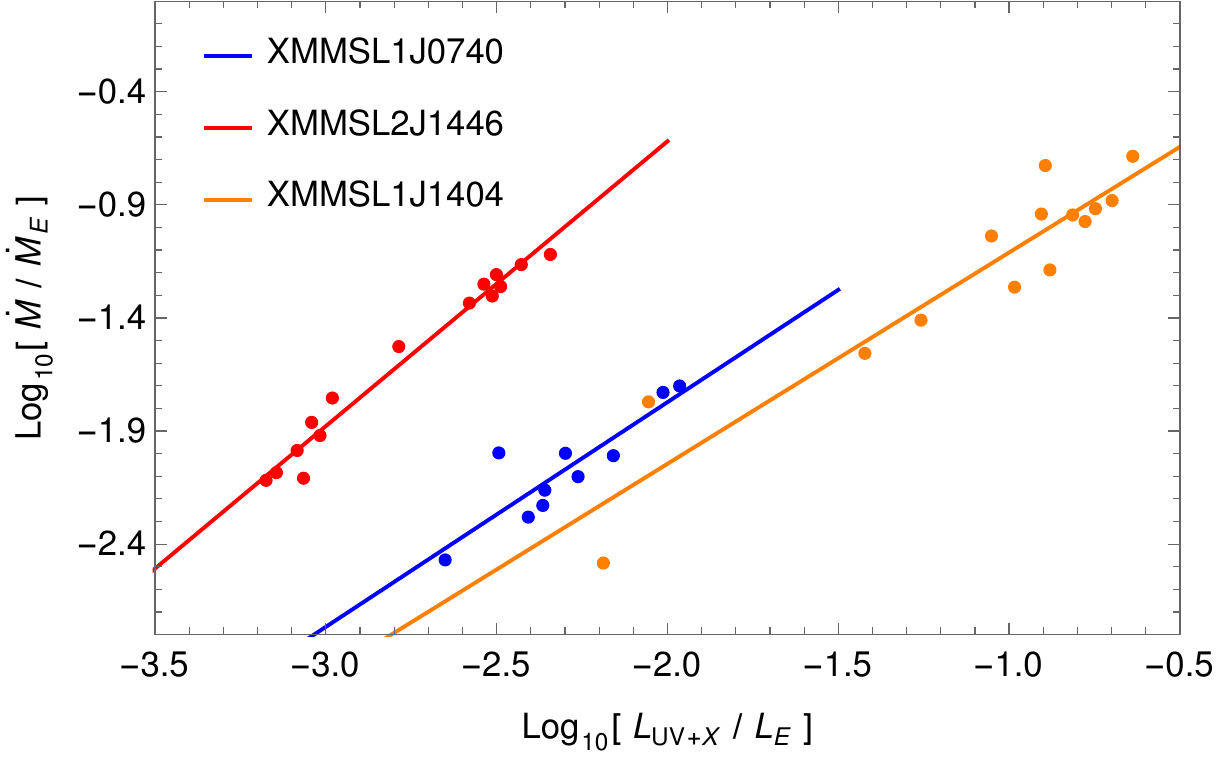}}
\subfigure[]{\includegraphics[scale=0.66]{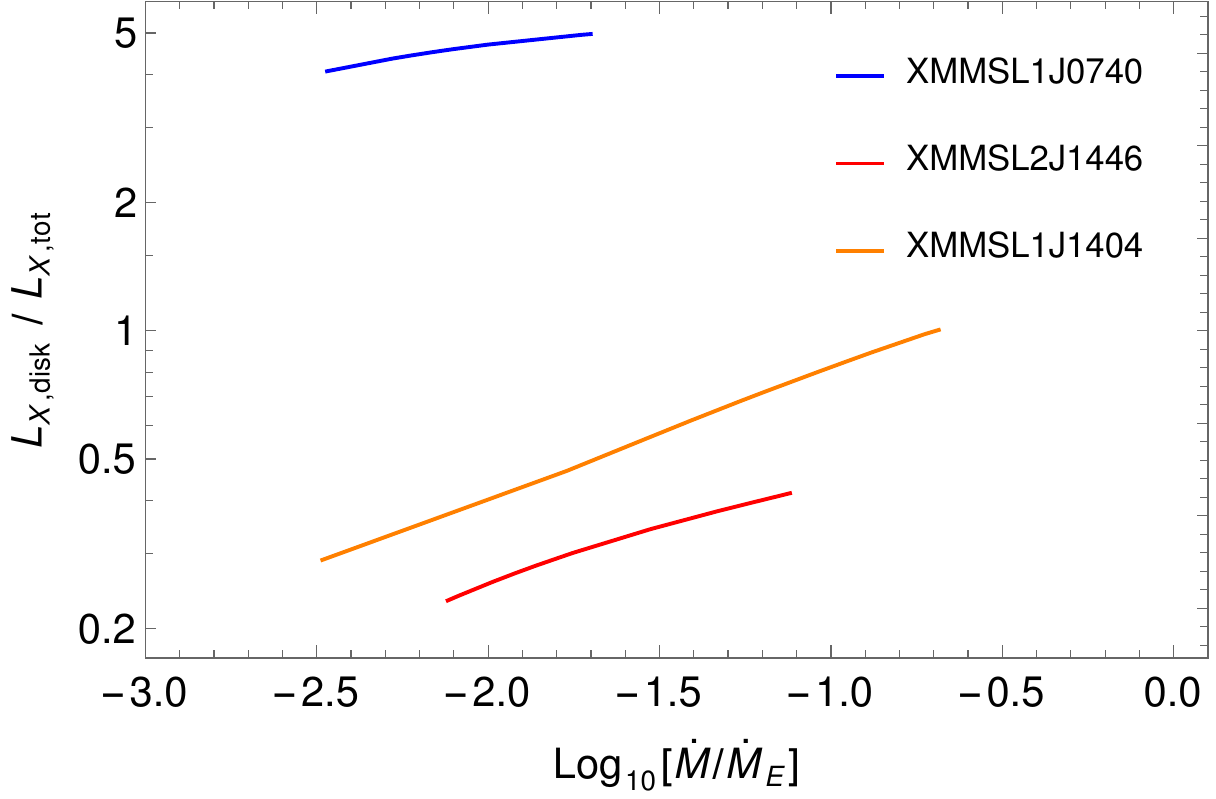}}
\end{center}
\caption{\label{nrmdot} (a) The estimated mass accretion rate corresponding to the individual observed luminosity points for the non-relativistic models by minimizing the equation (\ref{chi}) at individual points for the parameters given in Table \ref{nrres}. The solid lines are linear model fit given by $\log_{10}[\dot{M}/\dot{M}_E] = c_1 + u \log_{10}[(L_X+L_{\text{UV}})/L_E]$. The obtained values are $\{c_1,~u\}$ =  \{0.213, 0.993\} (blue), \{1.89, 1.26\} (red), \{-0.176, 0.934\} (orange). (b) The ratio of X-ray luminosity from the disk ($L_{\text{X,disk}}$) to corona ($L_{\text{X,tot}}$) as a function of obtained mass accretion rate. With an increase in $\dot{M}$, the X-ray luminosity from the disk increases and even dominates at a higher accretion rate.}
\end{figure*}

\noindent We assume the X-ray emission from the corona to be a power-law spectrum given by $ L_{\nu} =  K \nu^{1-\Gamma}$ with constant $K$ and Photon index $\Gamma$ in the range $\nu_{i} = 0.01$ keV and $\nu_f = 10~{\text{keV}}$ such that the total X-ray flux integrated over the spectrum is $\displaystyle{L_{\text{X,tot}} = K \int_{\nu_{i}}^{\nu_f} \nu^{1-\Gamma} \, \diff \nu}$. Then, the spectrum at any frequency is given by 

\begin{equation}
L_{\nu} = K L_{\text{2~keV}}^{1-\Gamma} =  L_{\text{X,tot}} (2-\Gamma) \frac{\nu^{1-\Gamma}}{\nu_f^{2-\Gamma}-\nu_i^{2-\Gamma}},
\end{equation}

\noindent and the luminosity is given by $\nu L_{\nu}$. FIG.~\ref{fitplt1} shows the model fit to the observations for the three TDE sources and the obtained parameters are given in Tables \ref{nrres} for the non-relativistic model. The $\mu$ is small for all the cases and implies that the viscous stress is dominated by the total pressure and the black hole has a low spin. The obtained mass accretion rate for the observed luminosity for three sources is shown in FIG.~\ref{nrmdot}a. The mass accretion rate for the source XMMSL1J1404 is close to the Eddington rate. We then fit the obtained mass accretion rate to the observed luminosity by a power-law relation and we found that $(L_X+L_{\text{UV}})/L_E \propto \dot{M}^{u_1}$, where $u_1 = 1/u = $ 0.794, 1.07 and 1.0069 for XMMSL2J1446, XMMSL1J1404 and XMMSL1J0740 respectively. The ratio of X-ray luminosity from the disk to X-ray luminosity from the corona given by $L_{\text{X,disk}}/L_{\text{X,corona}}$ increases with $\dot{M}$ that implies that the X-ray spectrum is soft for high mass accretion rate.

\begin{table*}
\caption{\label{nrres} The Parameter values obtained by minimizing the equation (\ref{chi}) using the non-relativistic disk-corona model along with the reduced chi square (see appendix \ref{ndcm}) for the considered TDE sources.}
\begin{ruledtabular}
\begin{tabular}{cccccccccc}
&&&&&&&&&\\
Sources & $\alpha_0$ & $\mu$ & $\eta$ & $a_d$ & $M_{\star}(M_{\odot})$ & $j$ & $q$ & $\chi_r^2$  \\
&&&&&&&&&\\
\hline
&&&&&&&&&\\
XMMSL1J0740 & 0.404 & $2.18 \times 10^{-7}$ & 0.779 & 0.01 & 0.91 & $3.9 \times 10^{-5}$ & 2.1 & 0.72  \\
&&&&&&&&&\\
\hline
&&&&&&&&&\\
XMMSL2J1446 & 0.238 & 0.033 & 0.324 & 0.331 & 1 & 0.78 & 4.45 & 1.35 \\
&&&&&&&&&\\
\hline
&&&&&&&&&\\
XMMSL1J1404 & 0.499 & 0.141 & 0.424 & 0.018 & 0.82 & 0 & 2.3 & 1.76 \\
&&&&&&&&&\\
\end{tabular}
\end{ruledtabular}
\end{table*}

\end{document}